\documentclass[11pt]{article}

\usepackage{jheppub}
\usepackage{graphicx,amsfonts, mathrsfs,amssymb}
\usepackage{amsmath,epsfig}
\usepackage{color}
\usepackage{latexsym}
\usepackage{pdfsync}
\usepackage{tikz}\usetikzlibrary{decorations.markings, arrows.meta}

\newbox\pippobox
\newcommand{\ld}{\lambda}
\newcommand{\p}{\partial}
\newcommand{\del}{\nabla}
\newcommand{\dta}{\delta}

\newcommand{\s}{\mathfrak{S}}
\newcommand{\V}{\mathfrak{V}}
\newcommand{\T}{\mathfrak{T}}
\newcommand{\la}{\langle}
\newcommand{\ra}{\rangle}

\title{Fluid/gravity correspondence: Second order transport coefficients in compactified D4-branes}
\author[a]{Chao Wu,\footnote{Stay at and become a member of the Wigner Research Center for Physics, Hungarian Academy of Sciences since November 5th 2016.}}
\author[a]{Yidian Chen}
\author[a,b]{and Mei Huang}
\affiliation[a]{Institute of High Energy Physics, Chinese Academy of Sciences, Beijing 100049, P.R. China}
\affiliation[b]{Theoretical Physics Center for Science Facilities, Chinese Academy of Sciences, Beijing 100049, P.R. China}

\emailAdd{wuchao@ihep.ac.cn}
\emailAdd{chenyd@ihep.ac.cn}
\emailAdd{huangm@ihep.ac.cn}

\abstract{We develop the boundary derivative expansion (BDE) formalism of fluid/gravity correspondence to nonconformal version through the compactified, near-extremal black D4-brane. We offer an explicit calculation of 9 second order transport coefficients, i.e., the $\tau_\pi,~\tau_\pi^*,~\tau_\Pi,~\lambda_{1,2,3}$ and $\xi_{1,2,3}$ for the strongly coupled, uncharged and nonconformal relativistic fluid which is the holographic dual of compactified, near extremal black D4-brane. We also show that the nonconformal fluid considered in this work is free of causal problem and admits the Haack-Yarom relation $4\lambda_1+\lambda_2=2\eta\tau_\pi$.}


\arxivnumber{1604.07765}

\begin{document}

\maketitle

\section{Introduction}

Relativistic hydrodynamics is an effective theory which deals well with dynamics of a large number of classical or quantum particles under the long wavelength, low frequency limit at nonzero temperature and/or chemical potential, and it has been very successfully used in describing phenomena for a wide scope of areas in high energy nuclear collisions, astrophysics as well as cosmology \cite{Rischke:1998fq,Rezzolla:rela_hydro_textbook}.

Fluid dynamics is described by the conservation of energy, momentum, and net charge number of the system, and the equations of motion (EOMs) are just the conservation equations of the conserved energy-momentum tensor $T_{\mu\nu}$ and conserved vector currents $J^\mu_a$. One has to input  initial conditions to uniquely solve these partial differential equations of the fluid dynamical EOMs. In the first-order hydrodynamical theories due to Eckart \cite{Eckart:relativistic_hydro} and Landau \cite{Landau:hydro_to_high_energy_collisions}, the conserved energy-momentum tensor and conserved currents are expanded by using the macroscopic degrees of freedom in the long wavelength and low frequency limit, i.e., the local energy density $\varepsilon$, the pressure density $p$, net density of charge $n_a$ of type $a$, 4-velocity $u^\mu$, metric $g_{\mu\nu}$ (in curved spacetime), and their gradients. As pointed out in Ref.\cite{Kranys:paradox_1st_order}, Eckart's theory has a severe problem that it admits infinite conducting speed of heat transfer which does not abide by the Einstein's principle of relativity. Ref.\cite{Kranys:paradox_1st_order}, on the basis of \cite{Vernotte:generalization_Fourier_Eq,Cattaneo:generalization_Fourier_Eq}, tries to fix this partly by generalizing the conducting equations of heat flow and temperature, and partly by changing the definition of heat flow 4-vector. Though Krany$\check{\rm s}$ goes one step further, this is not yet the final story. M\"uller \cite{Muller:2nd_relativistic_hydro} and later Israel and Stewart \cite{Israel:2nd_relativistic_hydro1,Israel:2nd_relativistic_hydro2,Israel:2nd_relativistic_hydro3} point out that the Krany$\check{\rm s}$'s work has problem in only considering first order viscous terms for the expression of entropy flux, which can be fixed by adding the second order viscous terms. Later investigations \cite{Hiscock&Lindblom1,Hiscock&Lindblom2} show that the second order theory of the M\"uller-Israel-Stewart type is the correct theory for relativistic dissipative hydrodynamics.

The strongest driven force for the development of relativistic hydrodynamics is the experiments in the Relativistic Heavy Ion Collider (RHIC), and Muronga is the first to apply the 2nd order relativistic hydrodynamics to RHIC physics \cite{Muronga:2nd_RHIC1,Muronga:2nd_RHIC2,Muronga:2nd_RHIC3,Muronga:2nd_RHIC4}. Strictly speaking, the M\"uller-Israel-Stewart theory is not the complete 2nd order theory. So theorists start the journey to search for the correct and complete 2nd order theory from both the weak coupling regime \cite{Betz:complete_2nd,Betz:2nd_kinetic_theory,
York:2nd_kinetic_thoery,Moore:conf_2nd,Moore:nonconf_2nd} and strong coupling regime \cite{Baier:AdSCFT_GreenKubo_2nd,
Arnold:AdSCFT_GreenKubo_2nd,Bhattacharyya:fluid/gravity,Bhattacharyya:entropy_flux,Bhattacharyya:AdS5_dilaton,
Banerjee:AdS5_U(1)charged,Erdmenger:AdS5_U(1)charged,vanRaam:2nd_AdS4,Haack:2nd_AdS(d+1),
Haack:2nd_AdS(d+1)&matter,Bhattacharyya:2nd_AdS(d+1),Grozdanov:2nd_univ_ident,Grozdanov:2nd_relation_violate,
Buchel:¦Ë_correction_to_¦Ç/s,Buchel:2nd_finite_coup1,Buchel:2nd_finite_coup2,Bigazzi:holo_nonconf_2nd_analytic,
Finazzo:holo_nonconf_2nd,Shaverin:lambdaGB_1st_hold,Shaverin:lambdaGB_2nd_violate}.

Before we look back on these literatures, we would like to offer the readers some general information on the second order relativistic fluid. In second-order theories of dissipative fluids, the space-time evolution of thermodynamic quantities are affected not only by the equation of state but also by dissipative, non-equilibrium processes. Thus the conservative energy-momentum tensor have to be expanded to include the dissipative quantities such as viscosity, thermal conductivity, diffusion and also the relaxation coefficients. Second-order theories is hyperbolic in structure, which lead to well-posed initial-value (Cauchy) problems, and also lead to causal propagation. Relaxation time is the distinguishing feature of second-order theories, and relaxation terms permit us to study the evolution of the dissipative fluxes. For an uncharged nonconformal relativistic fluid, its constitutive relation in Landau frame, in the most general form, can be formulated as
\begin{align}\label{Romatschke E-M tensor}
  T_{\mu\nu} &= pP_{\mu\nu} + \varepsilon u_\mu u_\nu - \left(2\eta\sigma_{\mu\nu} + \zeta P_{\mu\nu}\theta \right) + 2\eta\tau_\pi \left( \sideset{_\la}{}{\mathop D}\sigma_{\mu\nu\ra} + \frac13 \sigma_{\mu\nu}\theta \right) + 2\eta\tau_\pi^*\frac{\sigma_{\mu\nu}\theta}3 \cr
  & + \kappa \left( R_{\la\mu\nu\ra} - 2u^\rho u^\sigma R_{\rho\la\mu\nu\ra\sigma}\right) + \kappa^*2u^\rho u^\sigma R_{\rho\la\mu\nu\ra\sigma} \cr
  & + 4\ld_1\sigma_{\la\mu}^{~~\rho}\sigma_{\nu\ra\rho} + 2\ld_2\sigma_{\la\mu}^{~~\rho}\Omega_{\nu\ra\rho} + \ld_3\Omega_{\la\mu}^{~~\rho}\Omega_{\nu\ra\rho} + \ld_4\nabla_{\la\mu}\ln s\nabla_{\nu\ra}\ln s \cr
  & + P_{\mu\nu}(\zeta\tau_\Pi D\theta + 4\xi_1\sigma_{\rho\ld}\sigma^{\rho\ld} + \xi_2\theta^2 + \xi_3\Omega_{\rho\ld}\Omega^{\rho\ld} + \xi_4P^{\rho\ld}\nabla_\rho\ln s\nabla_\ld\ln s + \xi_5R \cr
  & + \xi_6u^\rho u^\ld R_{\rho\ld}),
\end{align}
where $P_{\mu\nu}=g_{\mu\nu}+u_\mu u_\nu$ is the spatial projection tensor, $\theta=\nabla_\rho u^\rho$ is the expansion viscous term and $R_{\mu\nu}$, $R_{\mu\nu\rho\sigma}$ are the Ricci tensor and Riemann tensor related with the metric $g_{\mu\nu}$. Here we use the same nomenclature for the second order transport coefficients as in Ref.\cite{Romatschke:2nd_stress_tensor}, which offers a standard prescription for constructing the energy-momentum tensor for uncharged relativistic fluid. The only difference from the conventions of Ref.\cite{Romatschke:2nd_stress_tensor} is that the shear tensor $\sigma_{\mu\nu}$ here is one half of that in Ref.\cite{Romatschke:2nd_stress_tensor}: $\sigma_{\mu\nu}=P_\mu^\rho P_\nu^\sigma\nabla_{(\rho}u_{\sigma)}-\frac13P_{\mu\nu}\nabla_\rho u^\rho$ with $\nabla_{(\rho}u_{\sigma)}=\frac12( \nabla_{\rho}u_{\sigma}+ \nabla_{\sigma}u_\rho)$. In order to keep balance, we put an additional factor of 2 in front of all the viscous terms that consist of the shear tensor. That is why the viscous terms of Eq.(\ref{Romatschke E-M tensor}) involving $\sigma_{\mu\nu}$ have additional factors of 2 or 4 compared with Ref.\cite{Romatschke:2nd_stress_tensor}. We also define the temporal or the comoving derivative ``$D$": $D = u^\mu\nabla_\mu$ and the spatial-projected traceless symmetrized tensor e.g.
\begin{align}
  A_{\la\mu\nu\ra} = P_\mu^\rho P_\nu^\ld A_{(\rho\ld)} - \frac13 P_{\mu\nu}P^{\rho\ld}A_{\rho\ld}.
\end{align}
With this definition one can see the shear viscous tensor is automatically spatial-projected traceless symmetrized tensor $\sigma_{\la\mu\nu\ra}=\nabla_{\la\mu}u_{\nu\ra}=\sigma_{\mu\nu}$. $\Omega_{\mu\nu}$ is the vorticity tensor and is defined as $\Omega_{\mu\nu}=P_\mu^\rho P_\nu^\ld\nabla_{[\rho}u_{\ld]}$ with $\nabla_{[\rho}u_{\ld]}=\frac12(\nabla_\rho u_\ld - \nabla_\ld u_\rho)$.

From Eq.(\ref{Romatschke E-M tensor}) we can learn that for a uncharged nonconformal relativistic fluid, one needs 2+15 transport coefficients to completely describe its dissipative properties up to second order. Among these coefficients, 2 of them: $\eta$ and $\zeta$ are the first order ones and the other 15 of them: $\tau_\pi,~\tau_\pi^*,~\tau_\Pi,~\kappa,~\kappa^*,~\ld_{1,\cdots,4}$ and $\xi_{1,\cdots,6}$ are the second order ones. But only 1+5 of the 2+15 coefficients will be left if the fluid is conformal and uncharged, which are $\eta$ in the 1st order and $\tau_\pi,~\kappa,~\ld_{1,2,3}$ in the second order. $\kappa,~\kappa^*,\xi_5$ and $\xi_6$ are related with curved metric thus will be vanish if the fluid is in Minkowski spacetime. This case is appropriate to the hot and dense plasma in heavy ion collisions.

For the listed references on fluid in the weak coupling, Refs.\cite{Betz:complete_2nd,Betz:2nd_kinetic_theory,York:2nd_kinetic_thoery} are in the dilute gas limit thus the method that the authors use is the kinetic theory via the Grad's moment expansion \cite{Grad:moment_expansion}. Refs.\cite{Moore:conf_2nd,Moore:nonconf_2nd} are in the continuum limit and the authors of these references employ the conventional linear response theory, in which the transport coefficients are calculated through the conventional Kubo formula \cite{Kubo:The_formula,Hosaya:Kubo_formula_field_theory} (for a modern pedagogical treatment of this subject, see e.g.
Ref.\cite{Kovtun:lecture}).

The above listed works for strongly coupled fluid are all via holography. Ref. \cite{Baier:AdSCFT_GreenKubo_2nd} directly calculates the 2nd order (in derivative expansion) 2-point correlated transport coefficients $\kappa$ and $\tau_\pi$ for the $\mathcal N=4$ SYM plasma via the Green-Kubo formalism \cite{Son:GreenKubo_formalism,Policastro:GreenKubo_shear,Policastro:GreenKubo_sound} of the fluid/gravity duality\footnote{The authors also find the 2nd order, three-point correlated coefficients $\lambda_1$ by comparing with the result of Ref. \cite{Heller:relax_time_AdSCFT}.}. Being aware of the original Green-Kubo formalism of fluid/gravity correspondence only gives the formulation for 2-point correlators, the authors of Ref.\cite{Barnes:3point_correlator_GreenKubo} generalize it to the case of 3-point correlators. Based on this,
Ref.\cite{Arnold:AdSCFT_GreenKubo_2nd} makes a direct calculation of the 2nd order, 3-point correlated transport coefficients: $\ld_1$, $\ld_2$ and $\ld_3$.

Refs.\cite{Bhattacharyya:fluid/gravity,Bhattacharyya:entropy_flux,Bhattacharyya:AdS5_dilaton,Banerjee:AdS5_U(1)charged,
Erdmenger:AdS5_U(1)charged,vanRaam:2nd_AdS4,Haack:2nd_AdS(d+1),Haack:2nd_AdS(d+1)&matter,
Bhattacharyya:2nd_AdS(d+1)} studies the 2nd order transport coefficients for the fluid of various situations in the framework of BDE formalism of fluid/gravity correspondence. Refs.\cite{Bhattacharyya:fluid/gravity,Bhattacharyya:entropy_flux,
Bhattacharyya:AdS5_dilaton,Banerjee:AdS5_U(1)charged,Erdmenger:AdS5_U(1)charged} investigate 4D relativistic fluid by using the asymptotic $AdS_5$ background, among which \cite{Bhattacharyya:fluid/gravity,Bhattacharyya:entropy_flux} set up the BDE formalism and study the 2nd order coefficients \cite{Bhattacharyya:fluid/gravity} and entropy flux \cite{Bhattacharyya:entropy_flux} of uncharged conformal fluid. Since the result of \cite{Bhattacharyya:fluid/gravity} is very representative, so we record it here in our conventions in the hope that it can help the readers to understand our result better.
\begin{align}\label{AdS5 2nd results}
  T_{\mu\nu}^{AdS_5} &= r_H^4 ( P_{\mu\nu} + 3 u_\mu u_\nu ) - r_H^3 \cdot 2\sigma_{\mu\nu} + r_H^2 \bigg[ \frac{2-\ln2}{2} \cdot 2\left( \sideset{_\la}{}{\mathop D}\sigma_{\mu\nu\ra} + \frac13 \sigma_{\mu\nu}\theta \right) \cr
  & + \left( R_{\la\mu\nu\ra} - 2u^\rho u^\sigma R_{\rho\la\mu\nu\ra\sigma} \right) + \frac12 \cdot 4\sigma_{\la\mu}^{~~\rho}\sigma_{\nu\ra\rho} - \ln2 \cdot 2\sigma_{\la\mu}^{~~\rho}\Omega_{\nu\ra\rho} \bigg], \cr
   \eta &= r_H^3,~~\eta\tau_\pi = \frac{2-\ln2}{2}r_H^2,~~\kappa=r_H^2,~~\ld_1=\frac12r_H^2,~~\ld_2 = - \ln2\cdot r_H^2,~~\ld_3=0.
\end{align}
Here we also add $\kappa=r_H^2$ from \cite{Baier:AdSCFT_GreenKubo_2nd} which is obtained by directly calculating the 2-point correlated Green-Kubo formula in $AdS_5$ black hole background. This result is also derived out in the Weyl-covariant formulation of BDE formalism \cite{Bhattacharyya:2nd_AdS(d+1)}. Thus the above equation is the complete summary for the constitutive relation of strongly coupled uncharged $\mathcal N=4$ SYM plasma corresponds to the $AdS_5$ black hole in the unit $1/2\kappa_5^{AdS_5}=1$. Note that from (\ref{AdS5 2nd results}) one can see that the Haack-Yarom relation $4\ld_1+\ld_2=2\eta\tau_\pi$ is satisfied, which will be discussed in detail later.

Different variations for the same system such as the presence of a dilaton dependent forcing term \cite{Bhattacharyya:AdS5_dilaton} and a U(1) conserved charge \cite{Banerjee:AdS5_U(1)charged,Erdmenger:AdS5_U(1)charged} are also investigated. Generalizations of the calculations of \cite{Bhattacharyya:fluid/gravity} to different dimensional asymptotic AdS spacetimes are also done, they are Ref.\cite{vanRaam:2nd_AdS4} in $AdS_4$ and Refs. \cite{Haack:2nd_AdS(d+1),Haack:2nd_AdS(d+1)&matter,
Bhattacharyya:2nd_AdS(d+1)} in $AdS_{d+1}$. Among these, Ref.\cite{Haack:2nd_AdS(d+1)&matter} adds matter fields in the
$AdS_{d+1}$ black hole background of \cite{Haack:2nd_AdS(d+1)} and Ref.\cite{Bhattacharyya:2nd_AdS(d+1)} generalizes it to the situation where the boundary is curved. $\ld$ corrections\footnote{$\ld$ here is the 't Hooft coupling.} to the 2nd transport coefficients of $\mathcal N=4$ SYM plasma are studied in Refs.\cite{Grozdanov:2nd_univ_ident,Buchel:¦Ë_correction_to_¦Ç/s,Buchel:2nd_finite_coup1,Buchel:2nd_finite_coup2}. The fluids corresponds to Gauss-Bonnet theories are studied in Refs.\cite{Shaverin:lambdaGB_1st_hold,Shaverin:lambdaGB_2nd_violate,Grozdanov:2nd_relation_violate}.

It is also interesting to talk about the classifications and constraints for the 2+15 transport coefficients here. One can make classifications in the following aspects \cite{Moore:nonconf_2nd}: 1) From perturbative field theory view point, the $\eta,~\zeta,~\tau_\pi,~\tau_\Pi,~\kappa,~\kappa^*$ and $\xi_{5,6}$ can be calculated from 2-point correlation function in the Green-Kubo formalism while $\tau_\pi^*,~\ld_{1,\cdots,4}$ and $\xi_{1,\cdots,4}$ are from the 3-point correlated function. This means in an effective action formalism, $\eta,~\zeta,~\tau_\pi,~\tau_\Pi,~\kappa,~\kappa^*$ and $\xi_{5,6}$ will be related with ``linear" terms while $\tau_\pi^*,~\ld_{1,\cdots,4}$ and $\xi_{1,\cdots,4}$ will be related with the ``nonlinear" terms in the effective Lagrangian. 2) From the relation with the flatness of spacetime, $\kappa,~\kappa^*$ and $\xi_{5,6}$ are the only 4 coefficients relate with curved metric. 3) From conformality, only $\eta,~\tau_\pi,~\kappa$ and $\ld_{1,2,3}$ will be present in a conformal fluid, appearance of any other coefficients except these will suggest the entry into the nonconformal regime. 4) From the view of being thermodynamical or dynamical, $\kappa,~\kappa^*,\ld_{3,4}$ and $\xi_{3,4,5,6}$ are the thermodynamical ones while $\eta,~\zeta,~\tau_\pi,~\tau_\pi^*,~\tau_\Pi,~\ld_{1,2}$ and $\xi_{1,2}$ are the dynamical ones. The reason for this can be found in Ref.\cite{Moore:nonconf_2nd}.

The above remarks are from Ref.\cite{Moore:nonconf_2nd} and we would like to add one observation here:
5) Ref.\cite{Bhattacharyya:2nd_constraint} shows that the 8 thermodynamical coefficients, i.e.  $\kappa,~\kappa^*,\ld_{3,4}$ and $\xi_{3,4,5,6}$ are constrained by positivity of the divergence of entropy flux and the number of the constraints is 5, while the dynamical sector is free of these constraints. So the independent 2nd order coefficients of any nonconformal uncharged relativistic fluid is 10. But why the number of the constraints is 5? A physical account can be found in Ref.\cite{Banerjee:constraints_from_partition_function} which is based on relating the constitutive relations of the relativistic fluid with the equilibrium partition function.

Let $\bar N^{(2)}_T,\bar N^{(2)}_V,\bar N^{(2)}_S$ and $\bar N^{(2)}_{S,tot}$ separately denote the number of tensors, vectors, scalars and scalars of total derivative which are composed of two partial derivatives acting on background fields. Such derivatives of background fields are the non-dissipative terms and the coefficients for such terms are the nondissipative ones. The non-dissipative terms will not disappear in the stationary equilibrium. We also define $\tilde N^{(2)}_T,\tilde N^{(2)}_V,\tilde N^{(2)}_S$ to stand for the number of tensors, vectors and scalars that are made of second derivative order of fluid variables such as $T,u^\mu,g_{\mu\nu}$, respectively. These terms are actually the terms that contributes the constituent relations of the fluid. In Landau frame, for example, $\tilde N^{(2)}_T+\tilde N^{(2)}_S$ viscous terms will appear in the viscous part of stress tensor and $\tilde N^{(2)}_V$ terms will be present in the dissipative part of vector current. Then, according to Ref.\cite{Banerjee:constraints_from_partition_function}, $\tilde N^{(2)}_T-\bar N^{(2)}_T+\tilde N^{(2)}_S-\bar N^{(2)}_S$ of the coefficients from stress tensor and $\tilde N^{(2)}_V-\bar N^{(2)}_V$ of the coefficients from the vector conserve current are the dissipative coefficients and will disappear when evaluated at stationary equilibrium.

For the case of nonconformal relativistic fluid without vector charge, one has $\tilde N^{(2)}_T=8,\tilde N^{(2)}_S=7$ for the viscous terms and $\bar N^{(2)}_T=4,\bar N^{(2)}_S=4,\bar N^{(2)}_{S,tot}=1$ for the background data. So the number of non-dissipative coefficients is $\bar N^{(2)}_T+\bar N^{(2)}_S=4+4=8$. In the scalar part of second derivative order background terms, only $\bar N^{(2)}_S-\bar N^{(2)}_{S,tot}=4-1=3$ will contribute to the partition function and there will also be 3 known coefficient functions of background fields appear with respect to each of those 3 scalar background terms. So these 3 known coefficient functions will help us to eliminate 3 out of the 8 relations which comes from equating the stress tensor made of background field data and the outcomes of the variations of partition function. Thus the relations left will be $8-3=5$, which are the number of constraints for the 8 non-dissipative coefficients.

The references that have been introduced above on the 2nd order transport coefficients of strongly coupled relativistic fluid are all in conformal situations. Kanitscheider et al. \cite{Kanitscheider:Dp_hydro} study the 1st order nonconformal hydrodynamics in Dp-branes in the framework of BDE formalism in Fefferman-Graham coordinate \cite{Gupta:BDE_FeffermanGraham}. They predict a rough form for the energy-momentum tensor of 2nd order but the explicit analytical results for the 2nd order transport coefficients are not given. Using this method, the authors of Ref.\cite{Bigazzi:holo_nonconf_2nd_analytic} offer the first analytic 2nd order transport coefficients for the nonconformal relativistic fluid corresponding to a scalar deformed $AdS_5$ black hole background. The first numerical calculation for the 2nd order transport coefficients of nonconformal fluid is done in
Ref.\cite{Finazzo:holo_nonconf_2nd} which builds upon an Einstein+Scalar bottom-up holographic model. The authors manage to plot numerically the temperature dependent behavior of $\tau_\pi,~\kappa,~\kappa^*,\ld_{3,4}$ and $\xi_{3,4,5,6}$ by making use of the Kubo relations derived out in Ref.\cite{Moore:nonconf_2nd} as well as the 5 constraints from
Ref.\cite{Bhattacharyya:2nd_constraint}. This numerical result is the first step towards nontrivial temperature dependence for the 2nd order transport properties of nonconformal fluid at strong coupling regime thus offers the crossover information for the quark-gluon plasma (QGP).

Though Refs.\cite{Bigazzi:holo_nonconf_2nd_analytic,Finazzo:holo_nonconf_2nd} have covered all the 2nd order transport coefficients for the uncharged nonconformal relativistic fluid, both of them do not jump out of the framework of $AdS_5$ black hole---their bulk spacetime are both deviations from $AdS_5$ black hole. In this work, we would like to offer a non-asymptotically $AdS_5$ background to holographically study the second order nonconformal relativistic fluid. With this purpose and based on our previous work \cite{Wu:fluid_gravity_D4_1st} where we generalize the BDE formalism of fluid/gravity correspondence \cite{Bhattacharyya:fluid/gravity} in compactified D4-brane at first order, we are going to move to the second order in the same background and calculate the transport coefficients. Through Ref.\cite{Wu:fluid_gravity_D4_1st} and this work, we want to offer a nonconformal counterpart to Bhattacharyya et al.'s $AdS_5$ construction for the BDE formalism \cite{Bhattacharyya:fluid/gravity} and improve our knowledge about the second order transport properties for nonconformal relativistic fluid.

This paper is organized in 7 sections. In this section we offer the readers some background knowledge and highlight our motivations.
The following sections begin with section 2, where we will give a very brief review on the technics of the fluid/gravity correspondence
in BDE formalism and results for the first order calculation for nonconformal fluid. Section 3 will be preliminaries of the second order calculation. In section 4, we will deal with the 2nd order constraint equations from the boundary fluid viewpoint and the results will be helpful when we investigate the dynamical equations in section 5 and express the constitutive relation in section 6. Then we use the results of 2nd order metric perturbations solved in section 5 to calculate the boundary stress tensor in section 6 and discuss the final result in section 7.

\section{Brief review of the first order}

In this section, we will review the setup of our framework at the first order very briefly in order to warm up for the second order. If the reader wants to learn more about it, we recommend her/him to Ref.\cite{Wu:fluid_gravity_D4_1st}, where we develop a nonconformal version of the fluid/gravity correspondence by using the compactified, near-extremal black D4-brane.

The complete action for compactified D4-brane in type IIA string theory in string frame is \cite{Bigazzi:dynamical_flavor_SS}
\begin{align}\label{10D total action string frame}
  S &= \frac1{2\kappa_{10}^2}\int d^{10}x \sqrt{-G^{(s)}} \left[ e^{-2\phi}\left(\mathcal R + 4(\nabla_{\hat M}\phi)^2 \right) - \frac{g_s^2}{2\cdot4!}F_4^2\right] \cr
  &- \frac1{\kappa_{10}^2} \int d^9x \sqrt{-H^{(s)}} e^{-2\phi}\mathcal K^{(s)} + \frac1{\kappa_{10}^2} \int d^9x \sqrt{-H^{(s)}} \frac5{2L}e^{-\frac73\phi},
\end{align}
where $L^3=\pi g_sN_cl_s^3$ and other details can be found in the appendix.
In the Einstein frame, the above action becomes
\begin{align}\label{10D total action Einstein frame}
  S &= \frac1{2\kappa_{10}^2} \int d^{10}x \sqrt{-G} \left[ \mathcal R - \frac12 (\del_{\hat M}\phi)^2 - \frac{g_s^2}{2\cdot4!} e^{\frac{\phi}2} F_4^2 \right] - \frac1{\kappa_{10}^2} \int d^9x \sqrt{-H} \mathcal K \cr
  &+ \frac1{\kappa_{10}^2} \int d^9x \sqrt{-H} \frac5{2L} e^{-\frac1{12}\phi}.
\end{align}
We use
\begin{align}\label{ansatz for dimensional reduction of 10D bulk metric}
  ds^2 = e^{-\frac{10}3A} g_{MN}dx^M dx^N + e^{2A+8B}dy^2 + L^2 e^{2A-2B}d\Omega_4^2
\end{align}
and
\begin{align}\label{ansatz for dimensional reduction of 9D boundary}
  ds^2 = e^{-\frac{10}3A} h_{MN}dx^M dx^N + e^{2A+8B}dy^2 + L^2 e^{2A-2B}d\Omega_4^2
\end{align}
to separately reduce the bulk and boundary part of the Einstein frame action (\ref{10D total action Einstein frame}). Here again, the details for the relate definitions and derivations can be found in the appendix.

Following the derivations in the appendix, the 5D reduced action is\footnote{Here we have set $L=1$.}
\begin{align}\label{action: total}
  S= S_{bulk} - \frac1{\kappa_5^2} \int d^4x \sqrt{-h} K + \frac1{\kappa_5^2} \int d^4x\sqrt{-h} \frac52 e^{-\frac53A-\frac1{12}\phi},
\end{align}
in which the second part of r.h.s. is the Gibbons-Hawking term and the third part is the counter term. $K$ is the trace of the external
curvature and $S_{bulk}$ is
\begin{align}\label{action: bulk}
  S_{bulk}&=\frac1{2\kappa_5^2}\int d^5x\sqrt{-g}\left[R-\frac12(\p\phi)^2-\frac{40}3(\p A)^2-20(\p B)^2-V(\phi,A,B)\right],\cr
  V&(\phi,A,B)=\frac92e^{\frac\phi2-\frac{34}3A+8B}-12e^{-\frac{16}3A+2B}.
\end{align}
Here $R$ is the Ricci scalar and $\phi,~A,~B$ are three scalar fields coupled with metric. This bulk action is first derived in Ref.\cite{Benincasa:hydro_SS} where Benincasa and Buchel derive the sound speed and the ratio $\zeta/\eta$ for the compactified black D4-brane background. The scalar field $\phi$ originates from the dilaton and $A,B$ characterise the radii of the $S^1$ and $S^4$ on which the original 10D compactified D4-brane background is reduced. The EOMs are also recorded here
\begin{align}
  E_{MN}&-T_{MN}=0,\label{EOM: Ein} \\
  \del^2\phi&-\frac{9}4e^{\frac\phi2-\frac{34}3A+8B}=0, \label{EOM: phi} \\
  \del^2A&+\frac{153}{80}e^{\frac\phi2-\frac{34}3A+8B}-\frac{12}5e^{-\frac{16}3A+2B}=0,\label{EOM: A} \\
  \del^2B&-\frac{9}{10}e^{\frac\phi2-\frac{34}3A+8B}+\frac35e^{-\frac{16}3A+2B}=0, \label{EOM: B}
\end{align}
where
\begin{align}
  E_{MN}\equiv R_{MN}-\frac12g_{MN}R
\end{align}
is the 5 dimensional Einstein tensor and
\begin{align}
  T_{MN}&\equiv \frac12\left(\p_M\phi\p_N\phi-\frac12g_{MN}(\p\phi)^2\right)+\frac{40}3\left(\p_M A\p_N A-\frac12g_{MN}(\p A)^2\right)\cr
  &+20\left(\p_M B\p_N B-\frac12g_{MN}(\p B)^2\right)-\frac12g_{MN}V
\end{align}
is the bulk energy-momentum tensor.

The compactified black D4-brane background can be written as
\begin{align}\label{Einstein frame Black D4}
  ds^2 &= H_4^{-\frac38}( -f(r)dt^2 + d\vec x^2 ) + H_4^{\frac58}\frac{dr^2}{f(r)} + H_4^{-\frac38}dy^2 + H_4^{\frac58}r^2 d\Omega_4^2, \cr
  e^\phi &= H_4^{-\frac14},~~F_4 = g_s^{-1}Q_4\epsilon_4,~~H_4 = 1 + \frac{r_{Q4}^3}{r^3},~~f(r) =1 - \frac{r_H^3}{r^3}.
\end{align}
In the above, $\phi$ is the dilaton field with zero vacuum value, $F_4$ is the Ramond-Ramond (RR) field magnetically coupled with the D4-brane, $\epsilon_4$ is the volume form on the unit 4-sphere, $g_s$ is the string coupling constant and $Q_4=(2\pi l_s)^3g_sN_c/\Omega_4$\footnote{$Q_4$ is defined through the normalization condition for $F_4$: $2\kappa^2\mu_4N_c=\int_{S^4}F_4$, where $2\kappa^2=(2\pi)^7l_s^8$ and $\mu_4=((2\pi)^4l_s^5)^{-1}$ is the D4-brane charge.}. The D4-branes lie in the directions of $\{x^i,y\}$ with $y$ a compact dimension of topology $S^1$ hence the name ``compactified black D4-brane". Note that $dy^2$ is written together with $d\Omega_4$ which is to address that it is also a compact direction as the 4-sphere. The near horizon limit of metric in Eq.(\ref{Einstein frame Black D4}) is
\begin{align}
  ds^2 &= \left(\frac rL\right)^\frac98( -f(r)dt^2 + d\vec x^2 ) + \left( \frac Lr \right)^\frac{15}8\frac{dr^2}{f(r)} + \left( \frac rL \right)^\frac98 dy^2 + L^\frac{15}8 r^\frac18 d\Omega_4^2, \label{D4 metric near horizon limit} \\
  e^\phi &= \left( \frac rL \right)^\frac34, \label{dilaton near horizon limit}
\end{align}
where $L^3=Q_4/3=\pi g_sN_cl_s^3$ and it is related with the Kaluza-Klein mass in the original framework of Sakai-Sugimoto model \cite{Sakai:SS_model} where the metric is a double Wick rotated version of Eq.(\ref{D4 metric near horizon limit}) in the directions of $t$ and $y$. $L$ and $r_H$ are the two parameters with dimension in this paper and all the physical results can be formulated in terms of them. In the following we will set $L=1$ thus only $r_H$ will appear in the physical results. One can restore the presence of $L$ if she/he wants to make the results look more reasonable in units.

The EOM can be solved by the following metric and scalar profiles
\begin{align}
  ds^2&=r^\frac53(-f(r)dt^2+d\vec x^2)+\frac{dr^2}{r^\frac43f(r)},~~f(r)= 1 - \frac{r_H^3}{r^3}\label{5d bulk metric}, \\
  e^\phi&=r^\frac34,~~e^A=r^\frac{13}{80},~~e^B=r^\frac1{10},\label{3 scalar profiles}
\end{align}
which is reduced from the background of compactified near-extremal black D4-brane. The metric (\ref{5d bulk metric}) is 5 dimensional asymptotically flat\footnote{This has been explained in \cite{Wu:fluid_gravity_D4_1st}. One can calculate the Ricci scalar and the square of the Rieman tensor which separately gives $R=-\frac{5(14r^3+r_H^3)}{6r^{11/3}}$ and $R_{MNPQ}R^{MNPQ}=\frac{25(62r^6+2r^3r_H^3+125r_H^6)}{108r^{22/3}}$, from which we can easily see that at $r\to0$, both of these two approach to zero. This confirms us that the metric (\ref{5d bulk metric}) is asymptotically flat.} and has a curvature singularity at $r=0$. The Hawking temperature of (\ref{5d bulk metric}) is $T=3r_H^{1/2}/(4\pi)$ which is also the temperature at thermal equilibrium of this system. Re-expressed (\ref{5d bulk metric}) in Eddington-Finkelstein coordinates $dv=dt+\frac{dr}{r^{3/2}f(r)}$ with the coordinates are boosted as $dv\to -u_\mu dx^\mu,~dx^i\to P^i_{~\mu}dx^\mu$, one will have
\begin{align}\label{5d boosted bulk metric}
  ds^2&= r^\frac53\left( -f(r)u_\mu u_\nu dx^\mu dx^\nu + P_{\mu\nu}dx^\mu dx^\nu \right) - 2r^\frac16u_\mu dx^\mu dr, \cr
  P_{\mu\nu}&=\eta_{\mu\nu}+u_\mu u_\nu,~~u^\mu=\gamma(1,\beta_i),~~\gamma=\frac1{\sqrt{1-\beta_i^2}}.
\end{align}
One can check that both (\ref{5d bulk metric}) and (\ref{5d boosted bulk metric}) are solutions of Einstein equation given that $u_\mu u^\mu=-1$. In the BDE formalism of fluid/gravity correspondence \cite{Bhattacharyya:fluid/gravity}, $r_H$ and $u^\mu$ in (\ref{5d boosted bulk metric}) are promoted to be $x^\mu$ dependent as $r_H\to r_H(x^\mu),~u^\mu\to u^\mu(x^\nu)$, which are called the collective modes. They capture the deviations from the thermo equilibrium of the bulk metric.

In the original formulation of the BDE formalism \cite{Bhattacharyya:fluid/gravity}, the boundary that the fluid lives in is flat. This is also true for the situation here. So the 2nd order viscous terms should not have those relate with $\kappa,~\kappa^*$ and $\xi_{5,6}$, since these four can only appear when the spacetime that the fluid resides in is curved. Then from the 5 constraints among the 2nd order thermodynamical transport coefficients \cite{Bhattacharyya:2nd_constraint}, neither should $\ld_4$ and $\xi_4$ be at present. To illustrate this, let us record the 5 constraints between the second order transport coefficients here from e.g. Ref.\cite{Moore:nonconf_2nd}:
\begin{align}
  \kappa^* =& \kappa - \frac T2 \frac{d\kappa}{dT}, \\
  \xi_5 =& \frac12\left( c_s^2T \frac{d\kappa}{dT} - c_s^2\kappa - \frac\kappa3 \right), \label{constraint:xi5}\\
  \xi_6 =& c_s^2 \left( 3T \frac{d\kappa}{dT} - 2T \frac{d\kappa^*}{dT} + 2\kappa^* - 3\kappa \right) - \kappa + \frac43\kappa^* + \frac{\ld_4}{c_s^2}, \label{constraint:xi6} \\
  \xi_3 =& \frac{3c_s^2T}2 \left( \frac{d\kappa^*}{dT}-\frac{d\kappa}{dT} \right) + \frac{3(c_s^2-1)}2(\kappa^*-\kappa) - \frac{\ld_4}{c_s^2} + \frac14 \left( c_s^2T \frac{d\ld_3}{dT} - 3c_s^2\ld_3 + \frac{\ld_3}3 \right), \label{constraint: xi3}\\
  \xi_4 =& -\frac{\ld_4}6 - \frac{c_s^2}2 \left( \ld_4+T\frac{d\ld_4}{dT} \right) + c_s^4(1-3c_s^2) \left( T \frac{d\kappa}{dT} - T \frac{d\kappa^*}{dT} + \kappa^* - \kappa \right) \cr
  &- c_s^6T^3 \frac{d^2}{dT^2}\left( \frac{\kappa-\kappa^*}{T} \right). \label{constraint: xi4}
\end{align}
Since the boundary fluid resides in a flat spacetime in the framework of fluid/gravity correspondence, so one has from Eqs.(\ref{constraint:xi6}), (\ref{constraint: xi3}) and (\ref{constraint: xi4}) that
\begin{align}
  \ld_4 &= 0, \\
  \xi_3 &= - \frac{\ld_4}{c_s^2} + \frac14 \left( c_s^2T \frac{d\ld_3}{dT} - 3c_s^2\ld_3 + \frac{\ld_3}3 \right), \\
  \xi_4 &= -\frac{\ld_4}6 - \frac{c_s^2}2 \left( \ld_4+T\frac{d\ld_4}{dT} \right) \label{xi4 relate lambda4},
\end{align}
considering $\kappa,\kappa^*,\xi_{5,6}$ are all 0. The above equations will further give
\begin{align}
  \ld_4 &= 0,~~~~ \xi_4 = 0, \\
  \xi_3 &= \frac14 \left( c_s^2T \frac{d\ld_3}{dT} - 3c_s^2\ld_3 + \frac{\ld_3}3 \right). \label{xi3 relate lambda3}
\end{align}
Thus in the original framework of fluid/gravity correspondence that the boundary metric is flat, $\ld_4$ and $\xi_4$ will always be 0, no matter whether the constitutive relation of the boundary fluid will have the viscous terms relate with $\ld_4$ and $\xi_4$. Said differently, the original BDE formalism as constructed in Ref.\cite{Bhattacharyya:fluid/gravity} can not capture the viscous information relate with $\ld_4$ and $\xi_4$ \footnote{The original framework of BDE formalism \cite{Bhattacharyya:fluid/gravity} has been generalize into a Weyl covariant form in Ref.\cite{Bhattacharyya:2nd_AdS(d+1)}, which makes the transport coefficients of curved spacetime like $\kappa$ reachable. But this Weyl covariant form is still in the conformal regime. For now we do not know whether there is a similar formulation for nonconformal backgrounds that can help us to extract the coefficients like $\kappa,\kappa^*$ etc. If the answer is positive, then one may get the information for $\ld_4$ and $\xi_4$ via such a generalized nonconformal BDE formalism.}.

From the above discussions one can see that the 6 out of 8 thermodynamical transport coefficients: $\kappa,~\kappa^*,~\xi_{5,6}$, $\ld_4$ and $\xi_4$ are actually out of reach in the original framework of the fluid/gravity correspondence. But $\ld_3$ and $\xi_3$ are reachable since they relate with the vorticity tensor and they will always be accompany with each other by (\ref{xi3 relate lambda3}). Thus we can make a pre-judgement that the potential candidates among the 15 second order coefficients that our work may derive out are $\tau_\pi,\tau_\pi^*,~\tau_\Pi,~\ld_{1,2,3}$ and $\xi_{1,2,3}$. It turns out that $\ld_3$ and $\xi_3$ are trivial which is like the case of $\ld_3$ in $\mathcal N=4$ SYM plasma \cite{Bhattacharyya:fluid/gravity,Arnold:AdSCFT_GreenKubo_2nd}.

We want to make a further discussion on the above 9 candidates here. Since $\tau_\pi$ and $\tau_\Pi$ are the relaxation time due to the dissipation caused by the flow of shear and expansion types, respectively. Thus these two should be present at this paper. $\tau_\pi^*$ is the indicator for the entry into nonconformal regime associated with $\tau_\pi$ so it should also be here. As for $\ld_{1,2,3}$ and $\xi_{1,2,3}$, the only message that we can confirm for now is $\xi_2$ should be here since it associates with $\theta^2$. For the rest we can only say if any (or both) of $\xi_{1,3}$ appear(s), the corresponding $\ld_{1,3}$ must also appear. Since viscous terms relate with $\xi_{1,3}$ are the corresponding ``trace" part of those relate with $\ld_{1,3}$.

The following steps are standard: 1. expand the boundary dependent metric with respect to derivatives of collective modes; 2. add perturbations into the expanded metric and solve them from the EOM of 5D bulk; 3. calculate the boundary stress tensor using the full 5D bulk metric with all the perturbations present and one can get the transport coefficients.

The full metric with first order perturbations in \cite{Wu:fluid_gravity_D4_1st} is
\begin{align}\label{1st order full metric}
  ds^2=&-r^\frac53\left(f(r_H(x),r)-\frac{F_k(r_H(x),r)}{r^3}\p_\rho u^\rho\right)u_\mu u_\nu dx^\mu dx^\nu
              -4r^\frac76u_\mu Du_\nu dx^\mu dx^\nu \cr
        &+r^\frac53F(r_H(x),r)\sigma_{\mu\nu}dx^\mu dx^\nu+r^\frac53\left(1+\frac13F(r_H(x),r)\p_\rho u^\rho\right)P_{\mu\nu}dx^\mu dx^\nu \cr
        &-2r^\frac16\left(1+F_j(r_H(x),r)\p_\rho u^\rho\right)u_\mu dx^\mu dr,
\end{align}
where $F(r),~F_j(r)$ and $F_k(r)$ are solved from 1st order dynamical equations for the metric perturbations:
\begin{align}
    F(r)=& \frac1{3\sqrt{r_H}}\bigg[ 2\sqrt3 \bigg( \arctan\frac{1-2\sqrt{r/r_H}}{\sqrt3} - \arctan
                 \frac{1+2\sqrt{r/r_H}}{\sqrt3} + \pi \bigg) \cr
            & + \ln\frac{(\sqrt r+\sqrt{r_H})^4(r+\sqrt{rr_H}+r_H)^2(r^2+rr_H+r_H^2)}{r^6}\bigg], \cr
    F_j(r) =& - \frac25\frac{r^\frac52-r_H^\frac52}{r^3-r_H^3}+\frac1{10}F(r), \cr
    F_k(r) =& \frac45r^\frac52-\frac15(r^3+2r_H^3)F(r).
\end{align}
One should note that the way we write functions like $F(r)$ etc. means the $r_H$ inside are $x$ independent while $F(r_H(x),r)$ means they depend on $x$.

As \cite{Kanitscheider:holography_for_nonconformal_brane,Skenderis:field_theory_limit_Dp-branes} have shown that one can build the precise holography correspondence for the nonconformal Dp-branes ($p\neq3$) just the same as for D3-brane in the so called ``dual frame", in which the near horizon limit of Dp-branes will have the topology of $AdS_{p+2}\times S^{8-p}$. For the compactified D4-brane case, the dual frame will be $AdS_5\times S^1\times S^4$. The only difference of the non-conformal brane in the dual frame from the D3-brane is the linear dilaton. Since the perturbation of dilaton is not turned on, the dilaton is just part of the background. Thus the framework of BDE formalism of fluid/gravity correspondence build for D3-brane can be applied to the cases of nonconformal Dp-branes. This justifies the nonconformal generalization of BDE formalism.

The boundary stress tensor of the 5D reduced gravity theory can be derived out from Eq.(\ref{action: total}) \cite{Wu:fluid_gravity_D4_1st} as
\begin{align}
  T_{\mu\nu} = \frac1{2\kappa_5^2} \lim_{r\to\infty}r^\frac53\cdot2 \left( K_{\mu\nu} - h_{\mu\nu}K - \frac52 r^{-\frac13} h_{\mu\nu} \right).
\end{align}
The 3 scalar fields will not contribute the boundary stress tensor since they depend only on $r$. The Hilbert-Einstein part together with the Gibbons-Hawking term of Eq.(\ref{action: total}) will give $(K_{\mu\nu}-h_{\mu\nu}K)$ with the derivations can be found in textbook. The counter term will contribute as $\frac52r^{-1/3}h_{\mu\nu}$, whose derivations can be found in Ref.\cite{Wu:fluid_gravity_D4_1st}. The stress tensor for the boundary relativistic fluid upto first order viscous terms is
\begin{align}\label{1st boundary stress tensor}
  T^{(0)+(1)}_{\mu\nu} = r_H^3\left( \frac12P_{\mu\nu} + \frac52u_\mu u_\nu \right) - r_H^\frac52 \left( 2\sigma_{\mu\nu}
  + \frac4{15}P_{\mu\nu}\p_\rho u^\rho \right).
\end{align}
In the above equation and the following, we set $2\kappa_5^2=1$. It will be restored at the end of this paper. $T^{(0)}_{\mu\nu}$ is the ideal fluid energy-momentum tensor which contains only the first two terms in the r.h.s. of (\ref{1st boundary stress tensor}).

\section{Setup of the second order calculation}

At every order of the BDE formalism, the first step is to get the correct expanded bulk metric. The second order calculation is much more complicated than the first order. In this section, we will give detailed accounts on how to expand the first order complete metric (\ref{1st order full metric}) to the second derivative order. Please note that we will not introduce the second order perturbations in this section which will be the main content of section 5.

Here we would also like to explain a little bit on the meaning of the ``second order" in both the title of this and the next sections. It means that these two sections deal with the physical information at the second derivative order (that is, the metrics, constraint equations and stress tensors with terms of two partial derivatives with respect to $x^\mu$). It does not refer to the order of the perturbations (in sections 3 and 4). To be more precise, we expand the 1st order full metric to the 2nd order in section 3 and derive out the constraint equations and Navier-Stokes equations of the 2nd derivative order in a purely fluid viewpoint in section 4. Both of these two sections are the preliminaries for solving the 2nd order perturbations in section 5.

We begin by expanding $r_H(x)$ and $\beta_i(x)$ to 2nd order in (\ref{1st order full metric}) as:
\begin{align}
  r_H(x)&=r_H+\dta r_H+\frac12\dta^2 r_H+\dta r_H^{(1)},~~\beta_i(x)=\dta\beta_i + \frac12\dta^2\beta_i,\cr
  u^\mu(x) &= \left( 1+\frac12\dta\beta_i\dta\beta_i \right)\dta^\mu_0 + \left( \dta\beta_j+\frac12\dta^2\beta_j \right)\dta^\mu_j,
\end{align}
where we denote $r_H\equiv r_H(0)$ and $\dta\#,~\dta^2\#$ are short for $x^\mu\p_\mu\#,~x^\mu x^\nu\p_\mu\p_\nu\#$ \footnote{\# stands for $r_H$ or $\beta_i$}. $r_H^{(1)}$ is the first order collective mode for the relativistic fluid on the boundary which is independent of the first order source $x^\mu\p_\mu r_H^{(0)}$ that comes from derivative expansion of $r_H^{(0)}$.\footnote{A point should be made clear that in general, the collective modes are expanded as $\#(x)=\#^{(0)}(x)+\#^{(1)}(x)+\cdots$, and we write $r_H^{(0)}(x)$ as $r_H(x)$ just for simplicity since only $r_H^{(0)}(x)$ and $r_H^{(1)}(x)$ are related with our discussions but not the full collective mode $r_H(x)$. $\beta_i^{(1)}$ can be set to zero, according to \cite{Bhattacharyya:fluid/gravity}.} Then $f(r_H(x),r)$, $F(r_H(x),r),~F_j(r_H(x),r)$ and $F_k(r_H(x),r)$ in (\ref{1st order full metric}) can be expanded as
\begin{align}
  f(r_H(x),r)&= f(r) - \frac{3r_H^2}{r^3}\dta r_H - \frac{3r_H^2}{2r^3}\dta^2r_H - \frac{3r_H^2}{r^3}\dta r_H^{(1)}
  -\frac{3r_H}{r^3}(\dta r_H)^2, \cr
  F(r_H(x),r)&= F(r) - \frac{F(r)+2rF'(r)}{2r_H}\dta r_H, \cr
  F_j(r_H(x),r)&= F_j(r) + \left( \frac{5r_H^\frac32r^3-6r_H^2r^\frac52+r_H^\frac92}{5(r^3-r_H^3)^2}-\frac{F(r)
  +2rF'(r)}{20r_H} \right) \dta r_H, \cr
  F_k(r_H(x),r)&= F_k(r) + \frac{(r^3-10r_H^3)F(r)+2r(r^3+2r_H^3)F'(r)}{10r_H}\dta r_H,
\end{align}
where $f(r),~F(r),~F_j(r)$ and $F_k(r)$ stand for functions with $r_H$ independent of $x$. To any one of them, e.g. $F(r)$, we may just denote it as $F$ and its derivatives as $F',~F''$ etc. in order to make the conventions simple. Thus (\ref{1st order full metric}) can be expanded to the second order with respect to boundary derivatives as
\begin{align}\label{input metric 2nd}
  ds^2=&\left[-r^\frac53f + \frac{3r_H^2}{r^\frac43}\dta r_H + \frac{3r_H^2}{2r^\frac43}\dta^2r_H + \frac{3r_H}{r^\frac43}(\dta
  r_H)^2 + \frac{3r_H^2}{r^\frac43}\dta r_H^{(1)} + \frac{r_H^3}{r^\frac43}\dta\beta_i\dta\beta_i
  - 4r^\frac76\dta\beta_i\p_0\beta_i\right. \cr
  & \left. + \frac{F_k}{r^\frac43}( \p\beta + \dta\p\beta + \dta\beta_i\p_0\beta_i )
  + \frac{(r^3 - 10r_H^3)F + 2r(r^3+2r_H^3)F'}{10r_Hr^\frac43}\dta r_H\p\beta \right]dv^2 \cr
  & + 2\left[ -\frac{r_H^3}{r^\frac43}\dta\beta_i - \frac{r_H^3}{2r^\frac43}\dta^2\beta_i - \frac{3r_H^2}{r^\frac43}\dta r_H\dta\beta_i - \frac{F_k}{r^\frac43}\dta\beta_i\p\beta + 2r^\frac76( \p_0\beta_i + \dta\p_0\beta_i+\dta\beta_j\p_j\beta_i ) \right.\cr
  &\left. - \frac{r^\frac53F}2 ( \dta\beta_j\p_i\beta_j + \dta\beta_j\p_j\beta_i ) \right]dx^idv + 2r^\frac16 \left[ 1 + \frac12\dta\beta_i\dta\beta_i + F_j(\p\beta + \dta\p\beta + \dta\beta_i\p_0\beta_i)\right. \cr
  &\left. + \left(\frac{5r_H^\frac32r^3 - 6r_H^2r^\frac52 + r_H^\frac92}{5(r^3-r_H^3)^2} - \frac{F + 2rF'}{20r_H} \right)\dta r_H\p\beta \right]dvdr \cr
  &+ \left[ r^\frac53\dta_{ij} + \frac{r_H^3}{r^\frac43}\dta\beta_i\dta\beta_j - 2r^\frac76( \dta\beta_i\p_0\beta_j
  + \dta\beta_j\p_0\beta_i ) + r^\frac53F(\p_{(i}\beta_{j)} + \dta\p_{(i}\beta_{j)} + \dta\beta_{(i}\p_{|0|}\beta_{j)})\right. \cr
  &\left.- \frac{r^\frac53(F+2rF')}{2r_H}\dta r_H\p_{(i}\beta_{j)}\right]dx^idx^j - 2r^\frac16\bigg( \dta\beta_i + \frac12\dta^2\beta_i
  + F_j\dta\beta_i\p\beta \bigg)dx^idr,
\end{align}
where $\p\beta\equiv\p_i\beta_i$, $\p_0\equiv\frac{\p}{\p v}$ and $\dta\beta_{(i}\p_{|0|}\beta_{j)}=\frac12(\dta\beta_i\p_0\beta_j
  + \dta\beta_j\p_0\beta_i)$.

\section{The second order constraints and Navier-Stokes equations}

In this section, we will discuss the constraint equation and the Navier-Stokes equation at the second derivative order. This section may look like a digression and not relate with the discussions in the previous section. Its significance lies in two folds: Firstly, section 5 will use the results of this section as a consistent check. Because this section derives out the constraint equations and Navier-Stokes equations of 2nd derivative order in a purely fluid point of view while section 5 is in the bulk gravity standpoint. Secondly, the derivation in section 5 and 6 sometimes need the results of this section.

The constraint equation is given by
\begin{align}\label{2nd constraint of T(0)}
  \p_\mu\p^\rho T^{(0)}_{\rho\nu}=0,
\end{align}
and the Navier-Stokes equation is given by
\begin{align}\label{Navier-Stokes of T(0+1)}
  \p^\mu T^{(0+1)}_{\mu\nu}=0.
\end{align}
From (\ref{2nd constraint of T(0)}), with $\mu,~\nu$ can be set to 0 or $i$, we can derive the following constraint relations satisfied by the second order spatial viscous terms:
\begin{align}
  \mathbf{s}_1+\frac25\mathbf{s}_2-\frac85\mathfrak S_1-\frac4{25}\mathfrak S_3=0, \label{2ndConstraintT(0)-scalar1}\\
  \mathbf{s}_2+\frac12\mathbf{s}_3-2\s_1+\frac2{15}\s_3-\frac12\s_4+\s_5=0, \label{2ndConstraintT(0)-scalar2}\\
  \mathbf{v}_{1i}+2\mathbf{v}_{2i}-\frac85\V_{1i}-\V_{2i}+2\V_{3i}=0, \label{2ndConstraintT(0)-vector1}\\
  \mathbf{v}_{1i}+\frac43\mathbf{v}_{4i}+\frac2{15}\mathbf{v}_{5i}-4\V_{1i}-\frac45\V_{2i}-\frac85\V_{3i}=0, \label{2ndConstraintT(0)-vector2}\\
  \mathbf{v}_{3i}+\frac7{15}\V_{4i}-\V_{5i}=0, \label{2ndConstraintT(0)-vector3}\\
  \mathbf{t}_{1ij}+2\mathbf{t}_{3ij}-4\T_{1ij}+\frac{14}{15}\T_{4ij}+\frac12\T_{5ij}+2\T_{6ij}=0. \label{2ndConstraintT(0)-tensor}
\end{align}
The explicit forms of this terms are given in Table \ref{tab:2nd viscous terms}.
\begin{table}
\centering
\begin{tabular}{|l|l|l|}
  \hline
  $~~\mathbf{1}$ of $SO(3)$ & ~~~ $\mathbf{3}$ of $SO(3)$ &~~~~~~~~ $\mathbf{5}$ of $SO(3)$ \\ \hline\hline
 $\mathbf{s}_1=\frac1{r_H}\p_0^2r_H$&$\mathbf{v}_{1i}=\frac1{r_H}\p_0\p_ir_H$&
 $\mathbf{t}_{1ij}=\frac1{r_H}\p_i\p_jr_H-\frac13\dta_{ij}\mathbf{s}_3$\\
  $\mathbf{s}_2=\p_0\p_i\beta_i$ & $\mathbf{v}_{2i}=\p_0^2\beta_i$ & $\mathbf{t}_{2ij}=\p_{(i}l_{j)}$ \\
  $\mathbf{s}_3=\frac1{r_H}\p_i^2r_H$ & $\mathbf{v}_{3i}=\p_0 l_i$ & $\mathbf{t}_{3ij}=\p_0\sigma_{ij}$ \\
  $\s_1=\p_0\beta_i\p_0\beta_i$ & $\mathbf{v}_{4i}=\frac95\p_j\sigma_{ij}-\p^2\beta_i$ & $\T_{1ij}=\p_0\beta_i\p_0\beta_j-\frac13\dta_{ij}\s_1$ \\
  $\s_2=l_i\p_0\beta_i$ & $\mathbf{v}_{5i}=\p^2\beta_i$ & $\T_{2ij}=l_{(i}\p_0\beta_{j)}-\frac13\dta_{ij}\s_2$ \\
  $\s_3=(\p_i\beta_i)^2$ & $\V_{1i}=\frac13\p_0\beta_i\p\beta$ & $\T_{3ij}=2\epsilon_{kl(i}\p_{j)}\beta_l\p_0\beta_k + \frac23\delta_{ij}\s_2$ \\
  $\s_4=l_il_i$ & $\V_{2i}=\epsilon_{ijk}\p_0\beta_j l_k$ & $\T_{4ij}=\sigma_{ij}\p\beta$ \\
  $\s_5=\sigma_{ij}\sigma_{ij}$ & $\V_{3i}=\sigma_{ij}\p_0\beta_j$ & $\T_{5ij}=l_il_j-\frac13\dta_{ij}\s_4$ \\
    & $\V_{4i}=l_i\p\beta$ & $\T_{6ij}=\sigma_{ik}\sigma_{kj}-\frac13\dta_{ij}\s_5$ \\
    & $\V_{5i}=\sigma_{ij}l_j$ & $\T_{7ij}=2\epsilon_{kl(i}\sigma_{j)l}l_k$ \\
\hline
\end{tabular}
\caption{\label{tab:2nd viscous terms}The list of all the second order derivatives of temperature and spatial velocity fields.}
\end{table}
$l_i=\epsilon_{ijk}\p_j\beta_k$ is the pseudo vector associates with the vorticity tensor $\Omega_{\mu\nu}$ and $\sigma_{ij}=\p_{(i}\beta_{j)}-\frac13\dta_{ij}\p\beta$ is the spatial components of $\sigma_{\mu\nu}$.

It is necessary to give some accounts for the meanings of the constraints, i.e. Eq.(\ref{2ndConstraintT(0)-scalar1}) to Eq.(\ref{2ndConstraintT(0)-tensor}). They come from expanding all the components of (\ref{2nd constraint of T(0)}) to second order, in which Eq.(\ref{2ndConstraintT(0)-scalar1}) and Eq.(\ref{2ndConstraintT(0)-scalar2}) are the $(00)$ and $(ii)$ components of (\ref{2nd constraint of T(0)}); (\ref{2ndConstraintT(0)-vector1}) and (\ref{2ndConstraintT(0)-vector2}) are the $(0i)$ and $(i0)$ components of (\ref{2nd constraint of T(0)}); the last two are the anti-symmetric and symmetric part of the traceless tensor sector of (\ref{2nd constraint of T(0)}), respectively.

The Navier-Stokes equations at the second order are got by expanding (\ref{Navier-Stokes of T(0+1)}) to second order and $\nu=0$ gives
\begin{align}
  \p^\mu T^{(0+1)}_{\mu0} = &- \frac{15}2r_H^2\p_0r_H^{(1)} + r_H^\frac52 \p_i\beta_j \p_i\beta_j + r_H^\frac52 \p_i\beta_j \p_j\beta_i - \frac25 r_H^\frac52 (\p\beta)^2 \cr
  & +r_H^3 x^\mu \bigg(-\frac{15}{2r_H} \p_\mu\p_0r_H - \frac{15}{r_H^2} \p_\mu r_H \p_0r_H - 6\p_\mu\beta_i\p_0\beta_i - \frac9{r_H} \p_\mu\beta_i \p_ir_H \cr
  & - \frac9{r_H} \p_\mu r_H \p\beta - 3 \p_\mu\p\beta \bigg).
\end{align}
One may expect to get an equation of $\p_0r_H^{(1)}$ with some 2nd order viscous terms in scalar sector according to \cite{Bhattacharyya:fluid/gravity}. So the offending $x$ dependent terms should be zero by itself, which can be shown by using the first order constraint equations \cite{Wu:fluid_gravity_D4_1st}:
\begin{align}
  \frac1{r_H} \p_0r_H = -\frac25 \p\beta, ~~~~\frac1{r_H} \p_ir_H=-2\p_0\beta_i.
\end{align}
Then the $x$ dependent part can be reexpressed as
\begin{align}
  &~v\bigg( -\frac{15}{2r_H}\p_0^2r_H - 3\p_0\p\beta + \frac65(\p\beta)^2 + 12\p_0\beta_i\p_0\beta_i \bigg)\cr
  & + x_i\bigg( -\frac{15}{2r_H}\p_0\p_ir_H - 3\p_i\p\beta + 6\p_0\beta_i\p\beta + 12\p_i\beta_j\p_0\beta_j \bigg).
\end{align}
One can check that the terms in the bracket behind $v$ is just Eq.(\ref{2ndConstraintT(0)-scalar1}) and terms in the bracket of $x_i$ is Eq.(\ref{2ndConstraintT(0)-vector2}). Thus $\nu=0$ component of Eq.(\ref{Navier-Stokes of T(0+1)}) finally gives one of the Navier-Stokes equations for the non-conformal hydrodynamics at the second order
\begin{align}\label{NavierStokes1}
  \frac1{r_H^{1/2}}\p_0r_H^{(1)}=\frac8{225}\s_3+\frac4{15}\s_5.
\end{align}

The $\nu=i$ component of (\ref{Navier-Stokes of T(0+1)}) is
\begin{eqnarray}
      && \p^\mu T^{(0+1)}_{\mu i} \cr
  & = & r_H^\frac52\bigg( \frac3{2r_H^{\frac12}} \p_ir_H^{(1)} - \frac35 \p_0\beta_j \p_i\beta_j - 2\p_0\beta_j \p_j\beta_i +\frac{11}{15} \p_0\beta_i\p\beta -\frac4{15} \p_i\p\beta - 2\p_j\sigma_{ij} + 10\p_0\beta_j\sigma_{ij} \bigg) \cr
  & + & r_H^3x^\mu \bigg( -\frac{18}5 \p_\mu\beta_i \p\beta + \frac3{r_H}\p_\mu r_H \p_0\beta_i + 3 \p_\mu\p_0\beta_i + \frac3{2r_H} \p_\mu\p_i r_H + 3 \p_\mu\beta_j \p_j\beta_i \cr
  & + & 3 \p_\mu\beta_i \p\beta \bigg),
\end{eqnarray}
where the $x$ dependent part can be rewritten as
\begin{align}
  &~v\bigg( 3\p_0^2\beta_i + \frac3{2r_H} \p_0\p_ir_H - \frac95 \p_0\beta_i \p\beta + 3 \p_0\beta_j \p_j\beta_i \bigg)\cr
  & + x_j \bigg( \frac3{2r_H} \p_j\p_ir_H + 3 \p_0\p_j\beta_i - 6 \p_0\beta_j\p_0\beta_i - \frac35 \p_j\beta_i \p\beta + 3  \p_j\beta_k \p_k\beta_i \bigg).
\end{align}
The brackets behind $v$ and $x_j$ are separately the Eq.(\ref{2ndConstraintT(0)-vector1}) and Eq.(\ref{2ndConstraintT(0)-tensor}), thus we have another Navier-Stokes equation at the second order as
\begin{align}\label{NavierStokes2}
  \frac1{r_H^{1/2}}\p_ir_H^{(1)}=\frac43\mathbf v_{4i} + \frac45\mathbf v_{5i} + \frac4{15}\V_{1i} - \frac7{15}\V_{2i}  - \frac{74}{15}\V_{3i}.
\end{align}
Equations (\ref{NavierStokes1}) and (\ref{NavierStokes2}) can be derived as the constraint equations from the bulk gravity theory, as will be shown in the next section.

\section{The second order perturbations}

We will solve the second order perturbations in this section. In considering the perturbations, we will adopt the scheme of \cite{Bhattacharyya:fluid/gravity} that we stick to the gauge that $g_{rr}=0$ and $g_{\mu r}\propto u_\mu$. This gauge offers us the convenience that one need not to consider the fluctuations of $(rr)$ and $(ir)$ components of the bulk metric. The covariant form for the full perturbation ansatz can be set as
\begin{align}\label{2nd perturbation ansatz covariant}
  ds_{(full~pert.)}^2 &= \frac{k(x,r)}{r^\frac43}u_\mu(x) u_\nu(x) dx^\mu dx^\nu + \frac{2r_H^3}{r^\frac43}P_\mu^\rho w_\rho(x,r) u_\nu(x) dx^\mu dx^\nu - 2r^\frac16 j(x,r) u_\mu(x) dx^\mu dr \cr
  &+ r^\frac53( \alpha_{\mu\nu}(x,r) + h(x,r)P_{\mu\nu} )dx^\mu dx^\nu.
\end{align}
The 2nd order of the perturbation ansatz is:
\begin{align}\label{2nd perturbation ansatz expanded}
  ds^2_{(2nd~order~pert.)}=\frac{k^{(2)}(r)}{r^\frac43}dv^2 - \frac{2r_H^3}{r^\frac43}w_i^{(2)}(r) dx^idv + 2 r^\frac16 j^{(2)}(r) dvdr + r^\frac53 \left( h^{(2)}(r) \dta_{ij} + \alpha^{(2)}_{ij}(r) \right) dx^idx^j.
\end{align}

The results for the second order perturbations will have the form like
\begin{align}
  \sum_I \mathcal F_I(r) \times \mathcal T_{I}^{2nd~viscous},
\end{align}
where $\mathcal F_I(r)$ are some functions of $r$ and $\mathcal T_{I}^{2nd~viscous}$ are the second order viscous terms listed in Table \ref{tab:2nd viscous terms}. We will begin to solve all these second order perturbations in the rest of this section, for the sake of simplicity, we set $r_H=1$ from now on and will restore it when giving our final result.

\subsection{The tensor part}

We begin with the tensor part as in the first order, the EOM for $\alpha_{ij}^{(2)}$ is
\begin{align}
  E_{ij}-\frac13\delta_{ij}\delta^{kl}E_{kl}=T_{ij}-\frac13\delta_{ij}\delta^{kl}T_{kl}.
\end{align}
When putting (\ref{input metric 2nd}) into the above equation, one gets the differential equation for $\alpha_{ij}^{(2)}$:
\begin{align}\label{diff eq of alpha(ij)}
  \frac{d}{dr}\bigg( r^4f \frac{d\alpha_{ij}^{(2)}}{dr} \bigg) &= \left( 6r - \frac52r^\frac32F - 2r^\frac52F' \right) (\mathbf t_{3ij}+\T_{1ij}) + \bigg[ \frac{10}3r - 15r^2FF_j + \frac52r^\frac32 \left( 2F_j + \frac7{15}F \right) \cr
  & - \frac{21}5r^\frac52 F' + 2r^3 \left( 4F_jF'+2FF'-\frac52FF_j' \right) - \frac12 \big( 2FF' + 4F'F_j - 4FF_j' + FF_k'  \cr
  &  - 2F_kF' \big) - \frac45r^\frac72 F'' + \frac16r^4 fF'^2 + r^4 \big( F'F_j' + 2F_jF'' + FF'' \big) \cr
  & + r \big( F'F_k' - 2F_jF'' - F'F_j' + F_kF'' - FF'' - FF_k'' \big) \bigg] \T_{4ij} + \left( 2r + \frac1{r^2} \right) \T_{5ij} \cr
  & + \left( 8r - 5r^\frac32 F + r^4 f F'^2 \right) \T_{6ij} + \left( r + \frac54 r^\frac32 F + r^\frac52 F' \right)\T_{7ij}.
\end{align}
One can see that the above second order differential equation is much more complex than its first order counterpart. So in general it will not have an analytical solution as in the first order case. Since we only care about its behavior at large $r$, thus we will take large $r$ expansion during the solving process. Another remark is that the l.h.s. of Eq.(\ref{diff eq of alpha(ij)}) is the same as the differential equation of $\alpha_{ij}^{(1)}$. This manifests the remarkable features of the BDE formalism of fluid/gravity duality: 1. this formalism is linear in $r$ direction and nonlinear in $x^\mu$ directions; and 2. the homogeneous part of the differential equations in the variable $r$ is the same at every order, but the nonhomogeneous part, i.e. the source part in the r.h.s. of every differential equations for the perturbation ansatz are different from order to order.

We write the solution of (\ref{diff eq of alpha(ij)}) formally as
\begin{align}
  \alpha_{ij}^{(2)} = \int_r^\infty \frac{-1}{x^4f(x)}dx \int_1^x S_{ij}^{(\alpha)}(y) dy,
\end{align}
where the source term $S_{ij}^{(\alpha)}$ is just the r.h.s. of (\ref{diff eq of alpha(ij)}). An important feature of $S_{ij}^{(\alpha)}(y)$ is that it has several independent branches. Because every second order viscous term can be seen as an independent branch and they can be solved independently. For example we may solve (\ref{diff eq of alpha(ij)}) with only $\mathbf t_{3ij}$ at present first, and then with only $\T_{4ij}$,... To get the final solution we need only add all these ``subsolutions" with only one viscous term present at a time together. But even does one solve it in this way, he/she still can not integrate (\ref{diff eq of alpha(ij)}) directly to get an analytic solution. Since we only care about its behavior at large $r$, thus we do the above integration in the following way: 1. calculate the inner integration directly; 2. expand the first integrated result with $\frac{-1}{x^4f(x)}$ multiplied at large $r$; 3. calculate the outer integration. The final result turns out to be
\begin{align}\label{2nd solution of alpha}
  \alpha_{ij}^{(2)} = & \bigg[ \bigg( \frac43 - \frac{\pi}{9\sqrt3} - \frac{\ln3}{3} \bigg)\frac1{r^3} \bigg] \big( \mathbf t_{3ij} + \T_{1ij} \big) + \bigg[ \frac8{3r} + \bigg( \frac{28}{45} - \frac{\pi}{45\sqrt3} - \frac{\ln3}{15} \bigg)\frac1{r^3} \bigg] \T_{4ij} - \frac1r \T_{5ij} \cr
   & + \bigg[ \frac4r + \frac4{3r^3} \bigg] \T_{6ij} + \bigg[ -\frac2r + \bigg( \frac{\pi}{18\sqrt3} + \frac{\ln3}6 \bigg) \frac1{r^3} \bigg] \T_{7ij}.
\end{align}
We keep the result to the order $1/r^3$ since only this order contributes to the boundary stress tensor. Since (\ref{2nd solution of alpha}) is already regular at $r=r_H$ and asymptotically to zero, thus all the integration constants for the solution of $\alpha^{(2)}_{ij}$ should be zero.

Here we would like to explain more about the integration constants. The vector perturbation $w_i$ is trivial at any order and it will not contribute to the boundary stress tensor so will be ignored in the discussion here.

The differential equations that $\alpha_{ij}^{(2)}$ and $h^{(2)}$ satisfy are both of second order in derivatives of $r$ and they have the form like:
\begin{align}
  \mathbf H(r) P^{(2)}(r) = S(r).
\end{align}
$P^{(2)}(r)$ stands for either $\alpha_{ij}^{(2)}$ or $h^{(2)}$ and $\mathbf H(r)$ is the second order differential operator for both of them. $S(r)$ is the source term for $P^{(2)}(r)$, it can be of single branch or multi branch. The branch refers to the spatial viscous tensors (presented in Table \ref{tab:2nd viscous terms}) that $S(r)$ contains. For example, the source $S_{ij}^{(\alpha)}(r)$ for $\alpha_{ij}^{(2)}$ is a sum of $\mathbf t_{3ij},\T_{1ij},\T_{4ij},\T_{5ij},\T_{6ij}$ and $\T_{7ij}$ with $r$ dependent coefficient functions, as can be seen in Eq.(\ref{diff eq of alpha(ij)}). So $\alpha_{ij}^{(2)}(r)$ has 5 branches\footnote{$S^{(\alpha)}_{ij}$ is a summation of 6 spatial viscous tensors, but the coefficient functions for $\mathbf t_{3ij}$ and $\T_{1ij}$ are the same thus can be treated equally as one branch.}. The differential equations for $j^{(2)}$ and $k^{(2)}$ are involved with $h^{(2)}$, it turns out that their equations are first order ones.

The differential equations are solved analytically by definite integrations. So we come across the problem for the integration constants here. For $\alpha_{ij}^{(2)}$ or $h^{(2)}$, there are two constants in every branch: one is fixed by the regularity at $r=r_H$, the other one is fixed by the normalization condition at the infinity. So in general we should write the solution for $\alpha_{ij}^{(2)}$ or $h^{(2)}$ as
\begin{align}
  P^{(2)} = \int_r^\infty \frac{-1}{x^4f(x)}dx \int_1^x S(y) dy + \sum_I \Bigg( \int_r^\infty \frac{-C^{(I)}_1}{x^4f(x)}dx + C^{(I)}_2 \Bigg),
\end{align}
where ``$I$" is summed over all the viscous terms appearing in the source. Since both of the source for $\alpha_{ij}^{(2)}$ and $h^{(2)}$ have 5 branches, both of them contain a total number of 10 integration constants. For $j^{(2)}$ and $k^{(2)}$, due to their equations are first order ones, both of them have 5 integration constants. The integration constants of $j^{(2)}$ are fixed by the normalization condition at $r\to\infty$ and those of $k^{(2)}$ are fixed by the restriction that the boundary stress tensor is in Landau frame. Here we offer Table \ref{Tab: integration constants} to make a summary on the integration constants.
\begin{table}[h]
  \centering
  \begin{tabular}{|c|c|c|c|}
    \hline
      & regularity at $r_H$ & normalization at $r\to\infty$ & requirement of Landau frame \\ \hline
    $\alpha^{(2)}_{ij}$ & 5 & 5 & 0 \\ \hline
    $h^{(2)}$ & 5 & 5 & 0 \\ \hline
    $j^{(2)}$ & 0 & 5 & 0 \\ \hline
    $k^{(2)}$ & 0 & 0 & 5 \\ \hline
  \end{tabular}
\caption{\label{Tab: integration constants} This table shows that the integration constants for the solutions of the 2nd order perturbations: rows stands for the perturbations while the columns represent the conditions used to fix the integration constants. The number of each cell tells how many integration constants are fixed in the corresponding condition for a certain sector of perturbations. The integration constants of the vector part need not to consider since the vector part does not contribute to the boundary stress tensor.}
\end{table}
It turns out that the integration constants for $\alpha_{ij}^{(2)}$, $h^{(2)}$ and $j^{(2)}$ are all zero and only the 5 of $k^{(2)}$ are non-trivial.

\subsection{The vector part}

The constraint equation of the vector part is
\begin{align}
  g^{r0}(E_{0i}-T_{0i})+g^{rr}(E_{ri}-T_{ri})=0.
\end{align}
Using the 2nd order expanded metric (\ref{input metric 2nd}), the above equation gives
\begin{align}\label{vector constraint}
  \frac1{r^{8/3}}\p_ir_H^{(1)}& = \bigg[ -\frac{20}{9r^{1/6}} + \bigg( \frac{100}9r^\frac13-\frac4{3r^{8/3}} \bigg)F_j
  + \frac{20F_k'}{9r^{5/3}} - \frac{10F_k}{9r^{8/3}} - \frac{10}9r^\frac43fF' \bigg] \mathbf v_{4i} \cr
  & + \bigg[ \frac4{9r^{1/6}} + \bigg(\frac{10}9r^\frac13-\frac4{9r^{8/3}}\bigg)F_j + \frac{2F_k'}{9r^{5/3}} - \frac{F_k}{9r^{8/3}} + \frac29r^\frac43fF' \bigg] \mathbf v_{5i} \cr
  & + \bigg[ \frac{16}{r^{1/6}} + \frac4{r^{37/6}f^2}\bigg(-\frac{12}5 + \frac2{5r^{5/2}} + r^\frac12 + 6r^3 - 5r^\frac72 \bigg) + \frac{4F'}3\bigg(\frac12r^\frac43f + \frac4{5r^{5/3}} + \frac{23}5r^\frac43 \bigg)\cr
  & - \frac43F''\bigg(\frac6{5r^{2/3}} + \frac{3r^\frac73}5\bigg) - \frac{12F}{5r^{8/3}} - \frac{4F_j}3\bigg(\frac{85}2r^\frac13 + \frac1{r^{8/3}}\bigg) + \frac{4F_j'}3\bigg(\frac6{r^{5/3}}-15r^\frac43\bigg) \cr
  & - \frac{19F_k}{3r^{8/3}} + \frac{2F_k'}{3r^{5/3}} - \frac{4F_k''}{r^{2/3}} \bigg] \V_{1i} + \bigg[ -\frac1{r^{1/6}} - \frac12r^\frac43fF' + \frac{2F_j}3\bigg(\frac52r^\frac13-\frac1{r^{8/3}}\bigg) + \frac{F_k'}{3r^{5/3}}\cr
  & - \frac{F_k}{6r^{8/3}} \bigg] \V_{2i} + \bigg[ -\frac6{r^{1/6}} - 3r^\frac43fF' + \frac{4F_j}3\bigg(\frac52r^\frac13 - \frac1{r^{8/3}}\bigg) + \frac{2F_k'}{3r^{5/3}} - \frac{F_k}{3r^{8/3}} \bigg] \V_{3i}
\end{align}
Expand the above equation to the order of $1/r^3$ one gets the vector constraint equation at the second order:
\begin{align}
  \frac1{r^{8/3}}\p_ir_H^{(1)} = \frac1{r^{8/3}}\bigg( \frac43\mathbf v_{4i} + \frac45\mathbf v_{5i} + \frac4{15}\V_{1i} - \frac7{15}\V_{2i}  - \frac{74}{15}\V_{3i} \bigg),
\end{align}
which is the second Navior-Stokes equation (\ref{NavierStokes2}).

The dynamical equation of the vector sector reads
\begin{align}
  E_{ri}-T_{ri}=0,
\end{align}
which gives
\begin{align}
  \frac{r^\frac12}2\frac{d}{dr} \bigg[ \frac1{r^2} \bigg( \frac{dw_i^{(2)}}{dr} \bigg) \bigg] = S^{(w)}_i.
\end{align}
Here $S^{(w)}_i$ is the source of vector perturbation.
\begin{align}
  S^{(w)}_i = & - \bigg(\frac5{3r^{3/2}} + \frac56F' - \frac{25}{6r}F_j + \frac53F_j'\bigg)\mathbf v_{4i}
                          + \bigg(\frac1{3r^{3/2}} + \frac16F' + \frac5{12r}F_j - \frac16F_j'\bigg)\mathbf v_{5i}\cr
  & + \bigg[ \frac1{r^{3/2}} + \frac3{2r^4f} + \frac{18}{r^9f^3} \bigg( \frac65r^\frac92-\frac{r^2}5-r^5 \bigg) + \frac5rF_j + F_j' + \frac{3F}{8r} - \frac{57}{10}F' - \frac{33}{10}rF''\bigg]\V_{1i} \cr
  & + \bigg( \frac5{4r^{3/2}} - \frac38F' + \frac5{8r}F_j - \frac14F_j' \bigg) \V_{2i} - \bigg( \frac1{2r^{3/2}} + \frac94F' - \frac5{4r}F_j + \frac12F_j' \bigg)\V_{3i}.
\end{align}
Three out of its five branches have the divergence part
\begin{align}
  S^{(w)}_i (r\to\infty) \to \frac{4\V_1+2\V_2+4\V_3}{r^{3/2}}.
\end{align}
So $w^{(2)}_i(r)$ can be integrated out as
\begin{align}
  w^{(2)}_i(r) =& - r^2(4\V_{1i} + 2\V_{2i} + 4\V_{3i}) + \int_r^\infty dx x^2 \int_x^\infty dy \frac2{y^\frac12}\bigg( S^{(w)}_i(y) - \frac{4\V_1+2\V_2+4\V_3}{y^{3/2}} \bigg) \cr
   =& \bigg(\frac{100}{63r^{1/2}} - \frac{37}{42r}\bigg)\mathbf v_{4i} + \bigg(\frac{46}{63r^{1/2}} - \frac{71}{210r}\bigg)\mathbf v_{5i} - \bigg( 4r^2 + \frac{26}{21r^{1/2}} - \frac3{2r} \bigg)\V_{1i} \cr
  & - \bigg( 2r^2 + \frac1{3r^{1/2}} - \frac{33}{280r} \bigg)\V_{2i} - \bigg( 4r^2 + \frac{86}{21r^{1/2}} - \frac{243}{140r}\bigg)\V_{3i},
\end{align}
where $- r^2(4\V_{1i} + 2\V_{2i} + 4\V_{3i})$ comes from the indefinite integral of the divergent part of the source term $S^{(w)}_i (r\to\infty)$. We record the result of the solution to vector perturbation for the convenience of the reader. In fact, the vector perturbation does't contribute to the boundary stress tensor. This dues to the fact that the vector part in our frame work is trivial, which is like the case in Ref.\cite{Bhattacharyya:fluid/gravity}. The perturbations can contribute to the boundary stress tensor only if they contain terms of order $1/r^3$.

\subsection{The scalar part}

The scalar sector is still the most complicated part at second order. The good news is that we will benefit a lot from the experiences that we required at solving it in the first order. The scalar part will contribute to the stress tensor of the fluid on the boundary. So we need to solve all the three scalar perturbations explicitly. Among the EOMs of $\phi$, $A$ and $B$, we only need to consider one of them since their EOMs will give the same differential equations in the situation that we do not turn on the perturbations for these 3 scalar fields.

The constraint equation of scalar sector is:
\begin{align}
  &g^{rr}(E_{r0}-T_{r0})+g^{r0}(E_{00}-T_{00})=0, \\
  &g^{rr}(E_{rr}-T_{rr})+g^{r0}(E_{r0}-T_{r0})=0.
\end{align}
The first one gives
\begin{align}\label{first scalar constraint}
  \frac1{r^{8/3}} \p_0 r_H^{(1)} =& \bigg[ \frac4{15r^{1/6}} + \frac4{5r^{19/6}} - \frac23r^\frac13fF_j - \frac{F_k}{3r^{8/3}} - \frac{F}{5r^{8/3}} + \frac2{15}r^\frac43fF' \bigg] \mathbf s_2 + \frac2{5r^{19/6}} \mathbf s_3 \cr
  & + \bigg[ -\frac8{5r^{19/6}} + \frac4{15r^{1/6}} - \frac{F}{5r^{8/3}} - \frac23r^\frac13fF_j - \frac{F_k}{3r^{8/3}} + \frac2{15}r^\frac43fF' \bigg] \s_1 \cr
  &+ \bigg[ \frac4{75r^6f^2} \bigg( 5r^\frac{10}3 - 6r^\frac{17}6 - 4r^\frac13 + \frac6{r^{1/6}} - \frac1{r^{8/3}} \bigg) + \frac8{75r^{19/6}} + \frac4{45r^{1/6}} - \frac{12}{75r^{3/8}}F  \cr
  &+ \frac6{75}r^\frac43F' + \frac2{45}r^\frac43fF' + \frac4{75}r^\frac73fF'' - \frac{2F_k}{5r^{8/3}} + \frac{2F_k'}{15r^{5/3}} \bigg] \s_3 - \frac2{5r^{19/6}} \s_4  \cr
  & + \bigg[ \frac4{15r^{1/6}} + \frac4{5r^{19/6}} + \frac2{15}r^\frac43fF' \bigg] \s_5.
\end{align}
After expanding to order $\mathcal O(1/r^4)$, the above equation gives:
\begin{align}
  \frac1{r^{8/3}} \p_0 r_H^{(1)} = \frac1{r^{8/3}}\bigg( \frac8{225}\s_3+\frac4{15}\s_5 \bigg) + \frac4{5r^{19/6}} \bigg( \mathbf{s}_2+\frac12\mathbf{s}_3-2\s_1+\frac2{15}\s_3-\frac12\s_4+\s_5 \bigg).
\end{align}
Note that terms in the second bracket in the above equation is Eq.(\ref{2ndConstraintT(0)-scalar2}) thus equals to 0. So we reproduce the first Navier-Stokes equation (\ref{NavierStokes1}). The second scalar constraint equation gives
\begin{align}\label{diff eq:scalar constraint}
       & 3(5r^3-2)h_{(2)}' - 30r^2j_{(2)} - 5k_{(2)}' \cr
   = & \bigg[ -8r + \frac52r^\frac32F + r^\frac52F' \bigg] \mathbf s_3 + \bigg[ 8r - 15r^\frac32F - 6r^\frac52F' \bigg] \s_1 + \bigg[ \frac{28}{15}r - \frac13r^\frac32F + 10r^\frac32F_j - 45r^2F_j^2 \cr
   & - \frac{74}{15}r^\frac52F' + \frac13(5r^3-2)FF' + (10r^3-4)F_jF' + 2F_kF' - \frac13r^4fF'^2 -10F_jF_k' + rF'F_k'  \cr
   & - \frac45r^\frac72F'' \bigg] \s_3 - \bigg[ \frac1{r^2} + 2r + \frac52r^\frac32F + r^\frac52F' \bigg] \s_4 + \bigg[ 2r + 5r^\frac32F + 2r^\frac52F' + (5r^3-2)FF' \cr
   & + \frac12r^4fF'^2 \bigg] \s_5.
\end{align}

There are 7 dynamical equations for the scalar part perturbations, they are the $(00)$, $(0r)$, $(rr)$ and $(ii)$ components of Einstein equation and the EOMs for those 3 scalar fields. But only 3 of them are actually independent, they are the $(rr)$ and $(ii)$ components of Einstein equation (\ref{EOM: Ein}) and the EOM of $\phi$ (\ref{EOM: phi})\footnote{The scalar constraints already make two of the $(00)$, $(0r)$ and $(rr)$ components of Einstein equation not independent, we choose the $(rr)$ component since it is the simplest. The EOMs of $\phi$, $A$ and $B$ give the same differential equations for the scalar part perturbations, so we choose the EOM of $\phi$.}. Here we will follow the procedure as what we did in solving the first order: we choose the $(rr)$ component of Einstein equation
\begin{align}\label{diff eq:Err-Trr}
  6rh_{(2)}'' + 9h'_{(2)} + 10j_{(2)}' =& \bigg( FF' + \frac r3F'^2 + \frac23rFF'' + 2rF'F_j' - 10F_jF_j' \bigg) \s_3 + \frac2{r^2}\s_4 \cr
  & + \big( rF'^2 + 2rFF'' + 3FF' \big)\s_5,
\end{align}
the EOM of $\phi$
\begin{align}\label{diff eq:phi}
      & r^3fj_{(2)}' + 6r^2j_{(2)} + k_{(2)}' - \frac32r^3fh_{(2)}' \cr
  = & \bigg[ r - \frac14r^\frac32F \bigg]\mathbf s_3 + \bigg[ 2r + \frac32r^\frac32F \bigg]\s_1 + \bigg[ -\frac25r - r^\frac32F_j + 9r^2F_j^2 + \frac1{30}r^\frac32F + \frac15r^\frac52F' \cr
     & - \frac16r^3fFF' - r^3fF'F_j + 3r^3fF_jF_j' + F_j'F_k - \frac12F'F_k + 2F_jF_k' \bigg]\s_3 + \frac14r^\frac32F\s_4 \cr
     & + \bigg[ \frac12FF' - \frac12r^\frac32F - \frac12r^3FF' \bigg]\s_5,
\end{align}
together with the second scalar constraint (\ref{diff eq:scalar constraint}) to solve the scalar perturbations. Eq.(\ref{diff eq:phi}) looks like a ``constraint equation" since only the first order derivative of the scalar perturbations are present. This is because we do not turn on the perturbation for $\phi$. If turned on, (\ref{diff eq:phi}) will be a second order differential equation for the perturbation of $\phi$ and, of course, with those first order derivative terms in (\ref{diff eq:phi}) at present, too.

Firstly, use (\ref{diff eq:scalar constraint}), (\ref{diff eq:Err-Trr}) and (\ref{diff eq:phi}) to remove $j$ and $k$, we get the differential equation for $h^{(2)}$:
\begin{align}\label{diff eq:h}
  \frac{d}{dr}\bigg( r^4f \frac{dh^{(2)}}{dr} \bigg) = S_h = c^{(h)}_1(r)\mathbf s_3 + c^{(h)}_2(r)\s_1 + c^{(h)}_{3}(r)\s_3 + c^{(h)}_4(r)\s_4 + c^{(h)}_5(r)\s_5,
\end{align}
where $S_h$ is the source term for the equation of $h^{(2)}$ and the coefficient functions $c^{(h)}$s are
\begin{align}
  c^{(h)}_1(r) &= - r + \frac{5}{12} r^{3/2} F + \frac{1}{3} r^{5/2} F',  \cr
  c^{(h)}_2(r) &= 6 r - \frac{5}{2}r^{3/2} F - 2 r^{5/2} F',  \cr
  c^{(h)}_3(r) &= - \frac{2r}{45} + \frac19r^4fFF'' + \frac13r^4fF'F_j' - \frac1{18}r^4fF'^2 - \frac53r^3fF_j F' - \frac19r^3fFF' + \frac{10}3r^3fF_jF_j'  \cr
                     & - \frac{4}{15}r^{7/2}F'' - \frac{59}{45} r^{5/2} F'+\frac{10}{3} r^3 F_j F'+\frac{5}{9} r^3 F F'-\frac{4}{3} F_j F' + \frac{1}{3} r F' F_k' - \frac{1}{6} F_k F' - \frac{2}{9}FF'  \cr
                      & - \frac1{18}r^{3/2}F + \frac53r^{3/2}F_j + \frac53F_kF_j', \cr
  c^{(h)}_4(r) &= -\frac{1}{3 r^2}-\frac{2 r}{3}+\frac{1}{3} r f-\frac{1}{3} r^{5/2} F'-\frac{5}{12} r^{3/2} F, \cr
  c^{(h)}_5(r) &= \frac{2 r}3 + \frac13 r^4 f F F'' + \frac13 r^4 f F'^2 - \frac13 r^3 f F F' + \frac23 r^\frac52 F' + \frac53 r^3FF' - \frac23FF' + \frac56 r^\frac32 F.
\end{align}
Thus $h^{(2)}$ can be solved by
\begin{align}
  h^{(2)}= \int_r^\infty \frac{-1}{x^4f(x)}dx \int_1^x S_h(y) dy.
\end{align}
This integral is done in the same way as we solve $\alpha^{(2)}_{ij}$, except for the order of expansion after finishing the inner integration. Here one should expand the result of the first integration to at least the order of $1/r^6$. This is because the 3 scalar perturbations are mixed together and they have different asymptotic behaviors:
\begin{align}
  F(r) &\simeq \frac4{r^\frac12} - \frac2{3r^3} + \mathcal O\left(\frac1{r^\frac72}\right), \label{asymptotic bahavior of F} \\
  F_j(r) &\simeq \frac1{3r^3} + \mathcal O\left(\frac1{r^\frac72}\right), \label{asymptotic bahavior of Fj} \\
  F_k(r) &\simeq \frac2{15} - \frac{12}{7r^\frac12} + \frac1{3r^3} + \mathcal O\left(\frac1{r^\frac72}\right). \label{asymptotic bahavior of Fk}
\end{align}
From the above one can see that the asymptotic behavior of $F$ is different from $F_j$ and $F_k$. That is to say terms of order $1/r^6$ in $F$ may still have effects on terms of order $1/r^3$ in $F_j$ and $F_k$ when solving the differential equations. Considering that terms of order $1/r^3$ will contribute to the boundary stress tensor, we should solve $h^{(2)}$ to the order of $1/r^6$ in order to get the right terms of order $1/r^3$ for $j^{(2)}$ and $k^{(2)}$. For the sake of simplicity, we will record the differential equations of $j^{(2)}$ and $k^{(2)}$ to the order of $1/r^6$ and the results of the 3 perturbations only to the order of $1/r^3$.

$h^{(2)}(r)$ can be integrated out directly as:
\begin{align}
  h^{(2)}(r) =& \bigg[\bigg( - \frac29 + \frac{\pi}{54\sqrt3} + \frac{\ln3}{18} \bigg)\frac1{r^3}\bigg]\mathbf s_3 + \bigg[\bigg( \frac43 - \frac{\pi}{9\sqrt3} - \frac{\ln3}3 \bigg) \frac1{r^3}\bigg]\s_1  \cr
   & + \bigg[ \frac4{9r} - \bigg( \frac{\pi}{405\sqrt3} + \frac{\ln3}{135} \bigg) \frac1{r^3} \bigg] \s_3 + \bigg[ \frac2{3r} + \bigg( \frac29 - \frac{\pi}{54\sqrt3} - \frac{\ln3}{18} \bigg) \frac1{r^3} \bigg] \s_4 \cr
   & + \bigg[ \frac4{3r} + \bigg( \frac{\pi}{27\sqrt3} + \frac{\ln3}9 \bigg) \frac1{r^3} \bigg] \s_5.
\end{align}
Since the above solution for $h^{(2)}$ is already regular at $r=r_H$ and asymptotically to zero at infinity. Thus all the integration constants for $h^{(2)}$ should be zero.

We would like to make a further explanation on the integration constants for the scalar sector. When we solve the first order \cite{Wu:fluid_gravity_D4_1st}, there is only one branch at present: $\p_i\beta_i$. The number of the integration constants is 4: two for $h$ and one for each of $j$ and $k$, because the differential equation for $h$ is second order in derivative of $r$ and first order for $j$ and $k$. The case is the same in the second order except that there are 5 branches at present now: $\mathbf s_3,~\s_1,~\s_3,~\s_4$ and $\s_5$. Thus the total number of integration constants should be $4\times5=20$ among which 10 of them belongs to $h^{(2)}$ while $j^{(2)}$ and $k^{(2)}$ separately has 5. For $h^{(2)}$, its 10 integration constants are all zero.

The differential equation for $j^{(2)}$ is
\begin{align}
  j_{(2)}'(r) = S_j = c^{(j)}_1(r)\mathbf s_3 + c^{(j)}_2(r)\s_1 + c^{(j)}_{3}(r)\s_3 + c^{(j)}_4(r)\s_4 + c^{(j)}_5(r)\s_5,
\end{align}
where $S_j$ is the source term for the equation of $j^{(2)}$ and the coefficient functions $c^{(j)}$s are
\begin{align}
  c^{(j)}_1(r)  =& \bigg( -1 + \frac{\pi}{12\sqrt3} + \frac{\ln3}4 \bigg)\frac1{r^4} + \frac7{5 r^{9/2}} -\frac{9}{10 r^5} +\bigg( - \frac{11}5 + \frac{11\pi}{60 \sqrt3} + \frac{11\ln3}{20} \bigg)\frac1{r^7},  \cr
  c^{(j)}_2(r)  =& \bigg( 6 - \frac{\pi}{2\sqrt3} - \frac{3\ln3}2 \bigg)\frac1{r^4} - \frac{42}{5 r^{9/2}} + \frac{27}{5 r^5} + \bigg( \frac{66}5 - \frac{11\pi}{10\sqrt3} - \frac{33\ln3}{10} \bigg)\frac1{r^7},  \cr
  c^{(j)}_3(r)  =& \frac2{15r^2} - \bigg( \frac{\pi}{90\sqrt3} + \frac{\ln3}{30} \bigg)\frac1{r^4} + \frac{28}{15r^{9/2}} - \frac{67}{25 r^5} - \bigg( \frac{286}{225} + \frac{11\pi}{450\sqrt3} + \frac{11\ln3}{150} \bigg)\frac1{r^7} \cr
                      & -\frac{1}{15} r F F''-\frac{1}{5} r F' F_j'-\frac{1}{30} r F'^2-\frac{1}{10} F F'+F_j F_j' , \cr
  c^{(j)}_4(r)  =& \bigg( 1 - \frac{\pi}{12\sqrt3} - \frac{\ln3}4 \bigg)\frac1{r^4} - \frac7{5 r^{9/2}} +\frac{9}{10 r^5} + \bigg( \frac{11}5 - \frac{11\pi}{60\sqrt3} - \frac{11\ln3}{20} \bigg)\frac1{r^7}, \cr
  c^{(j)}_5(r)  =& \frac2{5r^2} + \bigg( \frac{\pi}{6\sqrt3} + \frac{\ln3}2 \bigg)\frac1{r^4} - \frac{14}{5r^{9/2}} + \frac9{5r^5} + \bigg( \frac{22}{15} + \frac{11\pi }{30\sqrt3} + \frac{11\ln3}{10} \bigg)\frac1{r^7} \cr
                       & - \frac{1}{5} r F F''-\frac{1}{10} r F'^2-\frac{3}{10} F F'.
\end{align}

$j^{(2)}$ can be solved by
\begin{align}
  j^{(2)}(r) = - \int_r^\infty S_j(x) dx.
\end{align}
And the solution is
\begin{align}
  j^{(2)}(r) = & \bigg[\bigg( \frac13 - \frac{\pi}{36\sqrt3} - \frac{\ln3}{12} \bigg)\frac1{r^3}\bigg]\mathbf s_3 + \bigg[\bigg( -2 + \frac{\pi}{6\sqrt3} +\frac{\ln3}2 \bigg) \frac1{r^3}\bigg]\s_1 + \bigg[ \bigg( \frac{\pi}{270\sqrt3} + \frac{\ln3}{90} \bigg) \frac1{r^3} \bigg] \s_3 \cr
   & + \bigg[ \bigg( - \frac13 + \frac{\pi}{36\sqrt3} + \frac{\ln3}{12} \bigg) \frac1{r^3} \bigg] \s_4 + \bigg[ \bigg( \frac{\pi}{18\sqrt3} - \frac{\ln3}6 \bigg) \frac1{r^3} \bigg] \s_5.
\end{align}
From the large $r$ behavior of $F_j$ (\ref{asymptotic bahavior of Fj}) one can see that we still do not have to place integration constants here.

Put $h^{(2)}$ and $j^{(2)}$ into (\ref{diff eq:scalar constraint}), we gain the differential equation for $k^{(2)}$
\begin{align}
  k_{(2)}'(r) = S_k = c^{(k)}_1(r)\mathbf s_3 + c^{(k)}_2(r)\s_1 + c^{(k)}_{3}(r)\s_3 + c^{(k)}_4(r)\s_4 + c^{(k)}_5(r)\s_5,
\end{align}
where $S_k$ is the source term for the equation of $k^{(2)}$ and the coefficient functions $c^{(k)}$s are
\begin{align}
  c^{(k)}_1(r)  =& \frac{8r}5 + \frac1{15r^{3/2}} - \frac9{140r^2} + \bigg( - 1 + \frac{\pi}{12\sqrt3} + \frac{\ln3}4 \bigg)\frac1{r^4} + \frac{14}{15r^{9/2}} - \frac{18}{35r^5}  \cr
  & + \bigg( - \frac45 + \frac{\pi}{15\sqrt3} + \frac{\ln3}5 \bigg)\frac1{r^7} - \frac12 r^\frac32 F - \frac15 r^\frac52 F' , \cr
  c^{(k)}_2(r) =& - \frac{8r}5 - \frac2{5 r^{3/2}} + \frac{27}{70r^2} + \bigg( 6 - \frac{\pi}{2\sqrt3} - \frac{3\ln3}2 \bigg)\frac1{r^4} - \frac{28}{5r^{9/2}} + \frac{108}{35r^5}  \cr
                        & + \bigg( \frac{24}5 - \frac{6\ln3}5 - \frac{2\pi}{5\sqrt3}\bigg)\frac1{r^7} + 3 r^\frac32 F + \frac65 r^\frac52 F', \cr
  c^{(k)}_3(r) =& -\frac{128r}{75} + \frac{676}{315r^{3/2}} - \frac{73}{42 r^2} - \bigg( \frac{91}{225} + \frac{\pi}{90\sqrt3} + \frac{\ln3}{30} \bigg)\frac1{r^4} + \frac{56}{45r^{9/2}} - \frac{268}{175r^5}  \cr
   & - \bigg( \frac{104}{225} + \frac{2\pi}{225\sqrt3} + \frac{2\ln3}{75} \bigg)\frac1{r^7} + \frac1{15} r^\frac32 F - 2 r^\frac32 F_j + \frac{74}{75} r^\frac52 F' + \frac4{25}r^\frac72F'' + \frac1{15}r^4f F'^2  \cr
   &  + \frac2{15}FF' - \frac13r^3FF' + \frac45F_jF' - 2r^3F_jF' + 9r^2F_j^2 - \frac15rF'F_k' - \frac25F_kF' + 2F_jF_k', \cr
  c^{(k)}_4(r)  =& -\frac{8r}5  - \frac1{15r^{3/2}} + \frac{149}{140r^2} + \bigg( 1 - \frac{\pi}{12\sqrt3} - \frac{\ln3}4 \bigg)\frac1{r^4} - \frac{14}{15r^{9/2}} + \frac{18}{35r^5} \cr
  &  + \bigg( \frac45 - \frac{\pi}{15\sqrt3} - \frac{\ln3}5 \bigg)\frac1{r^7} + \frac12r^\frac32F + \frac15r^\frac52F', \cr
  c^{(k)}_5(r)  =& - \frac{22r}5 + \frac{170}{21r^{3/2}} - \frac{467}{70r^2} + \bigg( - \frac{14}{15} + \frac{\pi}{6\sqrt3} + \frac{\ln3}2 \bigg)\frac1{r^4} - \frac{28}{15r^{9/2}} + \frac{36}{35r^5}\cr
 & + \bigg( \frac8{15} + \frac{2\pi}{15\sqrt3} + \frac{2\ln3}5 \bigg)\frac1{r^7} - r^\frac32F - \frac25r^\frac52F'
 - \frac1{10}r^4fF'^2 + \frac25FF' - r^3FF'.
\end{align}
The asymptotic behavior of $S_k$ has a divergence term in the branch of $\s_1$:
\begin{align}
  S_k(r\to\infty) \to 8r\s_1,
\end{align}
which will contribute to the result in the form of indefinite integral $\int 8r dr=4r^2$. So $k^{(2)}$ can be solved by
\begin{align}\label{integral k}
  k^{(2)}(r) = 4r^2\s_1 - \int_r^\infty (S_k(x)-8x\s_1) dx + C_{k1}\mathbf s_3 + C_{k2}\s_1 + C_{k3}\s_3 + C_{k4}\s_4 + C_{k5}\s_5,
\end{align}
where $C_{k1}$ to $C_{k5}$ are the integration constants which can be fixed by requiring the boundary stress tensor in Landau frame:
\begin{align}
  C_{k1}&=\frac2{15}-\frac\pi{90\sqrt3}-\frac{\ln3}{30},~~C_{k2}=-\frac45+\frac{\pi}{15\sqrt3}+\frac{\ln3}5,
  ~~C_{k3}=\frac\pi{675\sqrt3}+\frac{\ln3}{225}, \cr
  C_{k4}&=-\frac{2}{15}+\frac\pi{90\sqrt3}+\frac{\ln3}{30},~~C_{k5}=-\frac\pi{45\sqrt3}-\frac{\ln3}{15}.
\end{align}
The reason for the existence of these integration constants can be figured out from the asymptotic behavior of $k^{(1)}$. From (\ref{asymptotic bahavior of Fk}), we can see that the lowest term of $F_k$ starts from constants, and one can not get constant terms from only the integration part $- \int_r^\infty (S_k(x)-8x\s_1) dx$ in (\ref{integral k}). So we need to add the $C_{k}$s ``by hand". Thus the final result for $k^{(2)}$ is
\begin{align}
  k^{(2)}(r)=&\bigg[\bigg(\frac2{15}-\frac\pi{90\sqrt3}-\frac{\ln3}{30}\bigg) - \frac1{20r} + \bigg(\frac13-\frac\pi{36\sqrt3}
  -\frac{\ln3}{12}\bigg)\frac1{r^3} \bigg]\mathbf s_3 \cr
  & + \bigg[4r^2 + \bigg(-\frac45+\frac{\pi}{15\sqrt3}+\frac{\ln3}5\bigg)+\frac3{10r}+\bigg(-2+\frac\pi{6\sqrt3}
  +\frac{\ln3}2\bigg)\frac1{r^3} \bigg]\s_1 \cr
  & + \bigg[ \bigg(\frac\pi{675\sqrt3} + \frac{\ln3}{225}\bigg) + \frac{16}{105r^{1/2}} + \frac{17}{10r} + \bigg(
  \frac2{45} + \frac\pi{270\sqrt3} + \frac{\ln3}{90} \bigg)\frac1{r^3} \bigg]\s_3 \cr
  & + \bigg[ \bigg( -\frac{2}{15}+\frac\pi{90\sqrt3}+\frac{\ln3}{30} \bigg) -\frac{19}{20r} + \bigg( -\frac13 + \frac\pi{36\sqrt3} + \frac{\ln3}{12} \bigg)\frac1{r^3} \bigg]\s_4 \cr
  & + \bigg[ \bigg(-\frac\pi{45\sqrt3}-\frac{\ln3}{15}\bigg) + \frac8{7r^{1/2}} + \frac9{10r} - \bigg( \frac\pi{18\sqrt3}
  + \frac{\ln3}6 \bigg)\frac1{r^3} \bigg]\s_5.
\end{align}

Collect all the 2nd order perturbations that we have solved in this section together with Eq.(\ref{input metric 2nd}), one can get the complete metric in global form up to second order:
\begin{align}\label{2nd global metric}
  ds^2=& - r^\frac53\left( f(r_H(x),r) - \frac{k(x,r)}{r^3} \right)u_\mu u_\nu dx^\mu dx^\nu - 2r^\frac76D(u_\mu u_\nu)dx^\mu dx^\nu \cr
 & + \frac{2r_H^3(x)}{r^\frac43}P_\mu^\rho w^{(2)}_\rho(x,r)u_\nu dx^\mu dx^\nu + r^\frac53( P_{\mu\nu} + h(x,r)P_{\mu\nu} + \alpha_{\mu\nu}(x,r)) dx^\mu dx^\nu \cr
         & - 2r^\frac16 (1+ j(x,r))u_\mu dx^\mu dr,
\end{align}
where $\#(x,r)=\#^{(1)}(x,r)+\#^{(2)}(x,r)~(\#=k,h,j,\alpha_{\mu\nu})$ and
\begin{align}
  k^{(1)}(x,r) &= F_k(r_H(x),r) \p_\rho u^\rho,~~h^{(1)}(x,r) = \frac13F(r_H(x),r)\p_\rho u^\rho,\cr
  j^{(1)}(x,r) &= F_j(r_H(x),r)\p_\rho u^\rho,~~\alpha_{\mu\nu}^{(1)}(x,r) = F(r_H(x),r)\sigma_{\mu\nu}.
\end{align}
The second order perturbations $\#^{(2)}(x,r)~(\#=k,h,j,\alpha_{\mu\nu})$ together with $w^{(2)}_\mu$ are of course taken their corresponding results solved in this section.

\section{The boundary stress tensor at the second order}

Much like the first order, the second order boundary stress tensor contains a tensor part and a scalar part which can be formally written as
\begin{align}
  T^{(2)}_{\mu\nu} = \pi^{(2)}_{\mu\nu} + P_{\mu\nu}\Pi^{(2)}.
\end{align}
The tensor part that we extract from the Brown-York energy-momentum tensor of the second order full metric is\footnote{Here we use ``$\sim$" because $r_H$ is not restored and $\kappa_5$ is not present.}
\begin{align}
  \pi^{(2)}_{ij} \sim& \bigg( 4-\frac\pi{3\sqrt3}-\ln3 \bigg)\bigg( t_{3ij}+\T_{1ij}+\frac{\T_{4ij}}3 \bigg) + \bigg( \frac85+\frac{2\pi}{15\sqrt3}+\frac{2\ln3}5 \bigg) \frac{\T_{4ij}}3 + 4\T_{6ij} \cr
                         & + \bigg( \frac\pi{6\sqrt3}+\frac{\ln3}2 \bigg)\T_{7ij},
\end{align}
and the scalar part is
\begin{align}
  \Pi^{(2)}\sim& -\bigg( \frac4{15}-\frac\pi{45\sqrt3}-\frac{\ln3}{15} \bigg)\mathbf s_3 + \bigg( \frac85-\frac{2\pi}{15\sqrt3}-\frac{2\ln3}5 \bigg)\s_1 - \bigg( \frac{2\pi}{675\sqrt3}+\frac{2\ln3}{225} \bigg)\s_3 \cr
   &+\bigg(\frac4{15}-\frac\pi{45\sqrt3}-\frac{\ln3}{15}\bigg)\s_4 + \bigg( \frac{2\pi}{45\sqrt3}+\frac{2\ln3}{15} \bigg)\s_5.
\end{align}
In order to get the covariant form of the boundary stress tensor, we use the replacement\footnote{There is a sign mistake for the replacement of $\T_{7ij}$ in the published version of this paper, which will cause $\ld_2$ to have an extra minus sign, thus the form of Haack-Yarom relation will also change.}
\begin{align}
  &t_{3ij}=\p_0\sigma_{ij} \to \sideset{_\la}{}{\mathop D}\p_\mu u_{\nu\ra},~~\T_{1ij} = \p_0\beta_i \p_0\beta_j - \frac13 \dta_{ij}\s_1 \to Du_{\la\mu}Du_{\nu\ra}, \cr
  &\T_{4ij}=\sigma_{ij}\p\beta\to\sigma_{\mu\nu}\p_\rho u^\rho,~~\T_{5ij}=l_il_j-\frac13\dta_{ij}\s_4\to l_\mu l_\nu-\frac13P_{\mu\nu}l_\rho l^\rho, \cr
  &\T_{6ij} = \sigma_{ik}\sigma_{kj}-\frac13\dta_{ij}\s_5\to \sigma_{\la\mu}^{~~\rho}\sigma_{\nu\ra\rho},~~
      \T_{7ij} = 2\epsilon_{kl(i}\sigma_{j)l}l_k \to -4\sigma_{\la\mu}^{~~\rho}\Omega_{\nu\ra\rho}
\end{align}
for the tensor sector and
\begin{align}
  &\mathbf s_3 = \frac1{r_H} \p_i^2r_H \to \frac1{r_H} P^{\mu\nu} \p_\mu\p_\nu r_H,~~\s_1=\p_0\beta_i\p_0\beta_i \to Du_\mu Du^\mu,~~\s_3 = (\p\beta)^2 \to (\p_\mu u^\mu)^2, \cr
  &\s_4=l_il_i\to l_\mu l^\mu=2\Omega_{\mu\nu}\Omega^{\mu\nu},~~\s_5=\sigma_{ij}\sigma_{ij}\to \sigma_{\mu\nu}
  \sigma^{\mu\nu}
\end{align}
for the scalar sector. Here we define $l_\mu=-\epsilon_{\mu}^{\;\nu\rho\sigma}u_\nu \partial_\rho u_\sigma $ with $\epsilon_{0123}=-\epsilon^{0123}=1$ and $\epsilon_{\lambda\mu\nu\rho}\epsilon^{\lambda\alpha\beta\gamma}=-\delta _{\mu\nu\rho}^{\alpha\beta\gamma}$. The scalar sector then becomes
\begin{align}\label{scalar part of stress tensor via direct replacement}
  \Pi^{(2)} &\sim - \bigg( \frac4{15} - \frac\pi{45\sqrt3} - \frac{\ln3}{15} \bigg) \frac1{r_H} P^{\mu\nu} \p_\mu\p_\nu r_H + \bigg( \frac85 - \frac{2\pi}{15\sqrt3} - \frac{2\ln3}5 \bigg) Du_\mu Du^\mu \cr
  & - \bigg( \frac{2\pi}{675\sqrt3} + \frac{2\ln3}{225} \bigg) (\p u)^2
  + \bigg(\frac4{15}-\frac\pi{45\sqrt3}-\frac{\ln3}{15}\bigg)2\Omega_{\mu\nu}^2 + \bigg( \frac{2\pi}{45\sqrt3}+\frac{2\ln3}{15} \bigg)\sigma_{\mu\nu}^2.
\end{align}
In order to match with the definition for the constituent relation of nonconformal fluid in \cite{Romatschke:2nd_stress_tensor}, we use the covariant form of (\ref{2ndConstraintT(0)-scalar2})
\begin{align}\label{covariant form of 2nd scalar constraint}
  D\p u=-\frac12 \frac1{r_H} P^{\mu\nu}\p_\mu\p_\nu r_H + 3Du_\mu Du^\mu - \frac2{15}(\p u)^2 + \Omega_{\mu\nu}^2 - \sigma_{\mu\nu}^2
\end{align}
to reexpress (\ref{scalar part of stress tensor via direct replacement}) as
\begin{align}\label{scalar part of stress tensor: standard form}
  \Pi^{(2)} \sim \bigg(\frac8{15}-\frac{2\pi}{45\sqrt3}-\frac{2\ln3}{15}\bigg) D\p u + \bigg( \frac{16}{225}-\frac{2\pi}{225\sqrt3}
   - \frac{2\ln3}{75} \bigg) (\p u)^2 + \frac8{15}\sigma_{\mu\nu}^2.
\end{align}
Thus the final form of boundary stress tensor upto second order derivative expansion is
\begin{align}\label{final result of stress tensor}
  T_{\mu\nu} =&\frac1{2\kappa_5^2}\Bigg\{ \frac12r_H^3P_{\mu\nu}+\frac52r_H^3u_\mu u_\nu - r_H^\frac52\bigg( 2\sigma_{\mu\nu} + \frac4{15}\p_\rho u^\rho P_{\mu\nu} \bigg) \cr
  & + r_H^2\bigg[ \bigg( 4 - \frac\pi{3\sqrt3} - \ln3 \bigg) \bigg( \sideset{_\la}{}{\mathop D}\sigma_{\mu\nu\ra} +\frac13
  \sigma_{\mu\nu}\p u \bigg) + \bigg( \frac85+\frac{2\pi}{15\sqrt3}+\frac{2\ln3}5 \bigg)\frac{\sigma_{\mu\nu}\p u}3\cr
  & + 4\sigma_{\la\mu}^{~~\rho}\sigma_{\nu\ra\rho} - \bigg( \frac{2\pi}{3\sqrt3} + 2\ln3\bigg ) \sigma_{\la\mu}^{~~\rho}
  \Omega_{\nu\ra\rho} \bigg] + r_H^2 P_{\mu\nu} \bigg[ \bigg( \frac8{15}-\frac{2\pi}{45\sqrt3}-\frac{2\ln3}{15} \bigg) D(\p u) \cr
  & + \bigg( \frac{16}{225} - \frac{2\pi}{225\sqrt3} - \frac{2\ln3}{75} \bigg)(\p u)^2 + \frac8{15}\sigma_{\mu\nu}^2 \bigg] \Bigg\}.
\end{align}
Here we restore $r_H$ and $\kappa_5$. Compare with the standard energy-momentum tensor of relativistic fluid, we can read all the 2nd order transport coefficients:
\begin{align}\label{2nd transport coefficients}
  & \eta\tau_\pi = \frac1{2\kappa_5^2}\bigg( 2 - \frac\pi{6\sqrt3} - \frac{\ln3}2 \bigg)r_H^2,~~\eta\tau_\pi^* = \frac1{2\kappa_5^2} \bigg( \frac45 + \frac{\pi}{15\sqrt3} + \frac{\ln3}5 \bigg)r_H^2,~~\lambda_1 = \frac1{2\kappa_5^2}r_H^2, \cr
  & \lambda_2 = - \frac1{2\kappa_5^2}\bigg( \frac{\pi}{3\sqrt3} + \ln3 \bigg)r_H^2,~~\ld_3=0,~~\zeta\tau_\Pi = \frac1{2\kappa_5^2}
  \bigg( \frac8{15} - \frac{2\pi}{45\sqrt3}-\frac{2\ln3}{15} \bigg)r_H^2, \cr
  & \xi_1=\frac1{2\kappa_5^2}\frac2{15}r_H^2,~~\xi_2 = \frac1{2\kappa_5^2} \bigg( \frac{16}{225} - \frac{2\pi}{225\sqrt3} - \frac{2\ln3}{75} \bigg) r_H^2,~~\xi_3=0.
\end{align}
The appearance of $\tau_\pi^*,~\tau_\Pi$ and $\xi_{1,2}$ indicates that we are in the nonconformal regime.

There are two simple relations among the 2nd order coefficients in (\ref{2nd transport coefficients}) given that $c_s^2=1/5$:
\begin{align}
  \tau_\pi = \tau_\Pi,~~~~\xi_1 = \frac{1-3c_s^2}3\ld_1 = \frac2{15}\ld_1.
\end{align}
These two relations match with the predictions that made in \cite{Romatschke:2nd_stress_tensor} about the nonconformal fluid of \cite{Kanitscheider:Dp_hydro}. And there are also relations that are not satisfied by our work, such as $\tau_\pi^*=-(1-3c_s^2)\tau_\pi$ and $\xi_2=\frac{1-3c_s^2}{3}2c_s^2\eta\tau_\pi$. But as it has been pointed out in Ref. \cite{Kleinert:universality_in_2nd_nonconformal_fluid} that both of these two relations miss $\ld_1$ and the authors suggest that the correct form of these two relations should be
\begin{align}
  \eta\tau^*_\pi = (1-3c_s^2) (4\ld_1-\eta\tau_\pi),~~~~\xi_2 = \frac29 (1-3c_s^2) [3c_s^2 \eta\tau_\pi + (1-6c_s^2)2\ld_1].
\end{align}
Using (\ref{2nd transport coefficients}) one can see that both of these two relations are also satisfied by our work.
%This may due to different asymptotic behaviors of the bulk metric in the first 4 directions, which in Ref.\cite{Romatschke:2nd_stress_tensor} is proportional to $r^2$ while in our case is $\sim r^\frac53$. This will cause the changes in the power law between thermal (and hydrodynamical) quantities such as entropy density (and transport coefficients) and temperature. For example the power law between entropy density and temperature is $s\sim T^3$ in \cite{Romatschke:2nd_stress_tensor} but it is $\sim T^5$ in our work. It is this deviations of the power law changes some relevant nonconformal transport coefficients thus violates some relations predicted in \cite{Romatschke:2nd_stress_tensor}.

The Haack-Yarom relation $4\ld_1+\ld_2=2\eta\tau_\pi$ is also satisfied by the coefficients in (\ref{2nd transport coefficients}). It is first found in Ref.\cite{Erdmenger:AdS5_U(1)charged} in charged $AdS_5$ black hole system to be satisfied for any value of chemical potential. The authors of \cite{Erdmenger:AdS5_U(1)charged} also point out that this relation is satisfied in asymptotic $AdS$ black holes of any dimension \cite{Haack:2nd_AdS(d+1)}. Later in Ref.\cite{Haack:2nd_AdS(d+1)&matter}, this relation is proved again to be satisfied in a large class of strongly coupled, conformal plasma of any dimension with matter fields \cite{Haack:2nd_AdS(d+1)&matter}. Further study \cite{Grozdanov:2nd_univ_ident} shows that it remains hold in $\mathcal N=4$ SYM plasma under $\alpha'^3\sim\ld^{-\frac32}$ string corrections. Even with the Gauss-Bonnet term added into the $AdS_5$ black hole background, this relation are shown to hold in the first order Gauss-Bonnet correction $\ld_{GB}$ \cite{Shaverin:lambdaGB_1st_hold}. But exception happens at the second order of $\ld_{GB}$ correction\footnote{We would like to thank S. Grozdanov for pointing this out for us.} which has been confirmed in both Refs. \cite{Shaverin:lambdaGB_2nd_violate} and \cite{Grozdanov:2nd_relation_violate}. The above results are all for the conformal relativistic fluid. For nonconformal case, Ref. \cite{Bigazzi:holo_nonconf_2nd_analytic} has shown that the validity of Haack-Yarom relation in some scalar field deformed asymptotic $AdS_5$ spacetime. Our result (\ref{2nd transport coefficients}) offers another solid confirmation for it. The Haack-Yarom relation is further proved to be held in some specific class of RG flows under the leading order nonconformal corrections \cite{Kleinert:universality_in_2nd_nonconformal_fluid}. In a word, if the manually added Gauss-Bonnet term of bulk gravity is not concerned, the Haack-Yarom relation $4\ld_1+\ld_2=2\eta\tau_\pi$ has a great possibility to be universal for both the conformal and nonconformal strongly coupled relativistic fluid.

The dispersion relation is got by working in the linearized regime of the fluid \cite{Bhattacharyya:fluid/gravity,Wu:fluid_gravity_D4_1st} and the results are:
\begin{align}\label{dispersion: comp D4}
  \omega_T(k) =& - \frac{i}{3r_H^\frac12}k^2 - \frac{i}{9r_H^\frac32}\bigg(2 - \frac\pi{6\sqrt3} - \frac{\ln3}2\bigg)k^4, \cr
  \omega_L(k) =& \pm\frac1{\sqrt5}k - i\frac4{15r_H^\frac12}k^2 \pm \frac{4\sqrt5}{75r_H}\bigg( \frac43 - \frac\pi{6\sqrt3} - \frac{\ln3}2 \bigg)k^3 \cr
                       & - i\frac{32}{225r_H^\frac32}\bigg( 2 - \frac\pi{6\sqrt3} - \frac{\ln3}2 \bigg)k^4.
\end{align}
where ``T'' and ``L" are short for ``transverse" and ``longitudinal", they represent for the shear and sound mode, respectively.

Grozdanov et al. have got the dispersion relations for the third derivative order relativistic fluid \cite{Grozdanov:3rd_hydro}. If we only count contributions of viscous tensors upto the second order, the dispersion relations for non-conformal fluid upto $k^4$ are:
\begin{align}\label{dispersion: general 2nd fluid}
  \omega_T(k) =& -i\frac\eta{\varepsilon+p}k^2 - i\frac{\eta^2\tau_\pi}{(\varepsilon+p)^2}k^4, \cr
  \omega_L(k) =& \pm c_sk - i\frac{ \frac23\eta+\frac12\zeta }{ \varepsilon+p }k^2 \pm \frac1{2c_s}\left[ \frac{ 2c_s^2\left( \frac23\eta\tau_\pi + \frac12\zeta\tau_\Pi \right) }{ \varepsilon+p } - \frac{ \left( \frac23\eta + \frac12\zeta \right)^2 }{ (\varepsilon+p)^2 } \right]k^3 \cr
  & - i\frac{ 2\left( \frac23\eta\tau_\pi + \frac12\zeta\tau_\Pi \right) \left( \frac23\eta + \frac12\zeta \right) }{ (\varepsilon+p)^2 }k^4.
\end{align}
Using the 1st (\ref{1st boundary stress tensor}) and 2nd (\ref{2nd transport coefficients}) order transport coefficients of this model, one can check that Eqs. (\ref{dispersion: comp D4}) and (\ref{dispersion: general 2nd fluid}) are consistent with each other.

At the end of this section, we would like to talk about the causality for the boundary fluid in this paper. According to \cite{Romatschke:2nd_stress_tensor,Romatschke:new_in_hydro}, a certain relativistic fluid respects causality when the group velocity of both the shear and sound modes are less than the speed of light (i.e., unity in natural units) in large $k$ limit. Using the related formulae in \cite{Romatschke:2nd_stress_tensor}, one can check that
\begin{align}
  \lim_{k\to\infty}\frac{d\omega_T}{dk} &= \sqrt{ \frac{\eta}{\tau_\pi(\varepsilon+p)} } \simeq 0.54 < 1, \\
  \lim_{k\to\infty}\frac{d\omega_L}{dk} &= \sqrt{c_s^2 + \frac43\frac{\eta}{\tau_\pi(\varepsilon+p)} + \frac{\zeta}{\tau_\Pi(\varepsilon+p)} } \simeq 0.82 < 1.
\end{align}
Thus the boundary relativistic fluid in our framework is causal.

\section{Discussions and outlooks}

We continue to investigate the 2nd order transport coefficients for the compactified, near-extremal black D4-brane in this paper based on our previous study \cite{Wu:fluid_gravity_D4_1st} via the BDE formalism of fluid/gravity duality \cite{Bhattacharyya:fluid/gravity}. We directly calculate 9 second order transport coefficients for the nonconformal relativistic fluid lives on the boundary. Our work successfully generalizes the BDE formalism into nonconformal background and offers a new set of directly and analytically calculated, 2nd order transport coefficients for strongly coupled, nonconformal relativistic fluid.

Here we want to compare the known transport coefficients between uncharged $AdS_5$ black hole and the compactified black D4-brane. The results are listed in Table \ref{tab:AdS5BH vs D4}. In the column of $AdS_5$ black hole, there are some coefficients belonging only to nonconformal fluid, we fill the blanks of such cases with a ``$\diagup$". Ref.\cite{Bhattacharyya:2nd_AdS(d+1)}, based on the construction of Refs.\cite{Loganayagam:Weyl_cova_formulation,Bhattacharyya:AdS5_dilaton}, reformulate the BDE formalism in the Weyl covariant language, which allows the boundary to be a curved spacetime but should belong to the same comformal class. This Weyl covariant version of BDE formalism can determine $\kappa$ for the conformal fluid, but we don't know whether a similar reformulation exist for the nonconformal fluid. So we just put a question mark for the compactified D4-brane. If the BDE formalism can be generalized to nonconformal fluid on the boundary, it will be possible to determine $\kappa,~\kappa^*$, $\xi_{5,6}$ and perhaps $\ld_4$ and $\xi_4$.

\begin{table}
\centering
\begin{tabular}{|c|c|c|}
  \hline
                            & $AdS_5$ black hole  & compactified black D4-brane \\ \hline
 $\eta$                 & $r_H^3$              & $r_H^{5/2}$ \\ \hline
 $\zeta$                & $\diagup$               & $\frac4{15}r_H^{5/2}$ \\ \hline
 $\eta\tau_\pi$     & $\frac{2-\ln2}{2}r_H^2$ & $\Big( 2 - \frac\pi{6\sqrt3} - \frac{\ln3}2 \Big)r_H^2$ \\ \hline
 $\eta\tau_\pi^*$ & $\diagup$ & $\Big( \frac45 + \frac{\pi}{15\sqrt3} + \frac{\ln3}5 \Big)r_H^2$ \\ \hline
 $\ld_1$               & $\frac12r_H^2$ & $r_H^2$ \\ \hline
 $\ld_2$               & $-\ln2\cdot r_H^2$ & $-\Big( \frac{\pi}{3\sqrt3} + \ln3 \Big)r_H^2$ \\ \hline
 $\ld_3$               & 0                       &  0  \\ \hline
 $\kappa$            &    $r_H^2$       & ? \\ \hline
 $\zeta\tau_\Pi$   & $\diagup$ & $\Big( \frac8{15} - \frac{2\pi}{45\sqrt3}-\frac{2\ln3}{15} \Big)r_H^2$ \\ \hline
 $\xi_1$               & $\diagup$ & $\frac2{15}r_H^2$ \\ \hline
 $\xi_2$               & $\diagup$ & $\Big( \frac{16}{225} - \frac{2\pi}{225\sqrt3} - \frac{2\ln3}{75} \Big) r_H^2$\\ \hline
 $\xi_3$               &  $\diagup$    &   0  \\ \hline
\end{tabular}
\caption{\label{tab:AdS5BH vs D4}A comparison of the known transport coefficients between $AdS_5$ black hole and compactified black D4-brane. The coefficients exist only in the nonconformal case will be marked with a ``$\diagup$" in the column of $AdS_5$ black hole. ``?" means that so far we don't know whether BDE formalism be capable to determine $\kappa$ in nonconformal case.}
\end{table}

In our final result of the 2nd order stress tensor (\ref{final result of stress tensor}), $r_H$ has been restored. But the dimension is still not correct. Since the dimensional parameter of the 5D bulk gravity are $\kappa_5,r_H$ and $L$. The result has already have $\kappa_5$ and $r_H$, thus we can make some repair on (\ref{final result of stress tensor}) in order to make its dimension correct. Through inserting $L$ to every term in the stress tensor, we will get the result under the full consideration of dimension as
\begin{align}\label{final result of stress tensor with right dimension}
  T_{\mu\nu} =&\frac1{2\kappa_5^2}\Bigg\{ \frac12 \frac{r_H^3}{L^4} P_{\mu\nu} + \frac52 \frac{r_H^3}{L^4} u_\mu u_\nu - \left( \frac{r_H}L \right)^\frac52 \bigg( 2\sigma_{\mu\nu} + \frac4{15}\p_\rho u^\rho P_{\mu\nu} \bigg) \cr
  & + \frac{r_H^2}L \bigg[ \bigg( 4 - \frac\pi{3\sqrt3} - \ln3 \bigg) \bigg( \sideset{_\la}{}{\mathop D}\sigma_{\mu\nu\ra} +\frac13
  \sigma_{\mu\nu}\p u \bigg) + \bigg( \frac85+\frac{2\pi}{15\sqrt3}+\frac{2\ln3}5 \bigg)\frac{\sigma_{\mu\nu}\p u}3\cr
  & + 4\sigma_{\la\mu}^{~~\rho}\sigma_{\nu\ra\rho} - \bigg( \frac{2\pi}{3\sqrt3} + 2\ln3\bigg ) \sigma_{\la\mu}^{~~\rho}
  \Omega_{\nu\ra\rho} \bigg] + \frac{r_H^2}L P_{\mu\nu}\bigg[ \bigg( \frac8{15} - \frac{2\pi}{45\sqrt3} - \frac{2\ln3}{15} \bigg) D(\p u) \cr
  & + \bigg( \frac{16}{225} - \frac{2\pi}{225\sqrt3} - \frac{2\ln3}{75} \bigg)(\p u)^2 + \frac8{15}\sigma_{\mu\nu}^2 \bigg] \Bigg\}.
\end{align}
As can be seen from the above, all terms in the constitutive relation has the dimension of $[\text{Mass}]^4$.

The stress tensor (\ref{final result of stress tensor with right dimension}) has already been in terms of 5D gravity language. In order to understand our result from the field theory side, here we would like to reformulate the result in terms of 4D field theory language. Note that our 5D gravity theory is equal to the 10D compactified near extremal D4-brane background, it is this 10D IIA string theory corresponds to the 4D field theory. In the string theory side, the parameters that we have are $r_H,g_s,l_s$ and $N_c$, they will relate with 5D field theory parameters directly by $g_5^2=(2\pi)^2g_sl_s$ and $\lambda_5=g_5^2N_cT_d$, where $g_5$ and $\ld_5$ are separately the 5D Yang-Mills and 't Hooft coupling, $T_d=3r_H^{1/2}/4\pi L^{3/2}$ is the deconfinement temperature for the 10D background \cite{Aharony:SS_deconfinement}. Note $N_c$ is also the field theory parameter. The 5D 't Hooft coupling relates with the 4D 't Hooft coupling by $\lambda=\lambda_5\beta_yT_d$. Following the way that \cite{Pang:SS_momentum_broadening} derives the entropy for 10D compactified near extremal D4-brane background in terms of field theory language, we can reformulate our result (\ref{2nd transport coefficients}) in terms of field theory quantities. For example, the energy and pressure density behave like $\sim r_H^3/(2\kappa_5^2L^4)$, and one can reformulate them in field theory language as
\begin{align}
  \frac1{2\kappa_5^2} \frac{r_H^3}{L^4} &= \frac{L^4\Omega_4\beta_y}{\kappa_{10}^2} \frac1{L^4} \left( \frac{4\pi}3 \right)^6 L^9 T^6 = \left( \frac{4\pi}3 \right)^6 \frac83\pi^2 \frac{\beta_y}{(2\pi)^7g_s^2l_s^8} (\pi g_s N_cl_s^3)^3 T^6 \cr
  &= \frac{2^6}{3^7} \pi^2 \cdot (2\pi)^2 g_s l_s \cdot N_c T_d \cdot \beta_y T_d \cdot N_c^2 \cdot \frac{T^6}{T_d^2} \cr
    & =  \left( \frac{4\pi}3 \right)^2 \frac{2^2}{3^5} \lambda N_c^2 \frac{T^6}{T_d^2}.
\end{align}
We summarize the result under full consideration of dimension in field theory language in Table \ref{tab:result of full dimension and field theory language}.

\begin{table}
\centering
\begin{tabular}{|c|c|c|}
  \hline
 $\varepsilon$           & $\frac1{2\kappa_5^2}\frac52\frac{r_H^3}{L^4}$ & $\frac52\left( \frac{4\pi}3 \right)^2 \frac{2^2}{3^5} \lambda N_c^2 \frac{T^6}{T_d^2}$ \\ \hline
 $p$                     & $\frac1{2\kappa_5^2}\frac12\frac{r_H^3}{L^4}$  & $\frac12\left( \frac{4\pi}3 \right)^2 \frac{2^2}{3^5} \lambda N_c^2 \frac{T^6}{T_d^2}$ \\ \hline
 $\eta$                 & $\frac1{2\kappa_5^2}\left(\frac{r_H}L\right)^\frac52$ & $\left( \frac{4\pi}3 \right)\frac{2^2}{3^5} \ld N_c^2 \frac{T^5}{T_d^2}$ \\ \hline
 $\zeta$                & $\frac1{2\kappa_5^2}\frac4{15}\left(\frac{r_H}L\right)^\frac52$ & $\frac4{15}\left( \frac{4\pi}3 \right)\frac{2^2}{3^5} \ld N_c^2 \frac{T^5}{T_d^2}$ \\ \hline
 $\eta\tau_\pi$     & $\frac1{2\kappa_5^2}\left(2-\frac\pi{6\sqrt3}-\frac{\ln3}2\right)\frac{r_H^2}L$ & $\left(2-\frac\pi{6\sqrt3}-\frac{\ln3}2\right)\frac{2^2}{3^5} \ld N_c^2 \frac{T^4}{T_d^2}$ \\ \hline
 $\eta\tau_\pi^*$ & $\frac1{2\kappa_5^2}\left( \frac45 + \frac{\pi}{15\sqrt3}+\frac{\ln3}5 \right)\frac{r_H^2}L$ & $\left( \frac45 + \frac{\pi}{15\sqrt3}+\frac{\ln3}5 \right)\frac{2^2}{3^5} \ld N_c^2 \frac{T^4}{T_d^2}$ \\ \hline
 $\ld_1$               & $\frac1{2\kappa_5^2}\frac{r_H^2}L$ & $\frac{2^2}{3^5} \ld N_c^2 \frac{T^4}{T_d^2}$ \\ \hline
 $\ld_2$               & $-\frac1{2\kappa_5^2}\left(\frac{\pi}{3\sqrt3}+\ln3\right)\frac{r_H^2}L$ & $-\left(\frac{\pi}{3\sqrt3}+\ln3\right)\frac{2^2}{3^5} \ld N_c^2 \frac{T^4}{T_d^2}$ \\ \hline
 $\ld_3$               &  0  &  0 \\ \hline
 $\zeta\tau_\Pi$   & $\frac1{2\kappa_5^2}\left(\frac8{15}-\frac{2\pi}{45\sqrt3}-\frac{2\ln3}{15}\right)\frac{r_H^2}L$ & $\left(\frac8{15}-\frac{2\pi}{45\sqrt3}-\frac{2\ln3}{15}\right)\frac{2^2}{3^5} \ld N_c^2 \frac{T^4}{T_d^2}$ \\ \hline
 $\xi_1$              & $\frac1{2\kappa_5^2}\frac2{15}\frac{r_H^2}L$ & $\frac2{15}\frac{2^2}{3^5} \ld N_c^2 \frac{T^4}{T_d^2}$ \\ \hline
 $\xi_2$               & $\frac1{2\kappa_5^2}\left(\frac{16}{225}-\frac{2\pi}{225\sqrt3}-\frac{2\ln3}{75}\right)\frac{r_H^2}L$ & $\left(\frac{16}{225}-\frac{2\pi}{225\sqrt3}-\frac{2\ln3}{75}\right)\frac{2^2}{3^5} \ld N_c^2 \frac{T^4}{T_d^2}$ \\ \hline
 $\xi_3$               &   0  &  0  \\ \hline
\end{tabular}
\caption{\label{tab:result of full dimension and field theory language} Reformulation of the result under full consideration of dimension in field theory language.}
\end{table}

Our result covers the whole sector of dynamical 2nd order transport coefficients. These coefficients satisfy the Haack-Yarom relation
and some other relations proposed in Refs. \cite{Romatschke:2nd_stress_tensor,Kleinert:universality_in_2nd_nonconformal_fluid}.
Comparing with Ref. \cite{Bhattacharyya:fluid/gravity}, we derive 5 more second order coefficients: $\tau_\pi^*,\tau_\Pi$ and $\xi_{1,2,3}$ that indicate the non-conformality. $\ld_3$ and $\xi_3$ are still zero in this work, similar as the case of $\ld_3$ in Refs.\cite{Arnold:AdSCFT_GreenKubo_2nd,Bhattacharyya:fluid/gravity}.

If one wants to study the transport properties for orders higher than two, he/she should begin with the 2nd order complete metric of global form (\ref{2nd global metric}) and expand it to the 3rd order in boundary derivatives just as the procedure of 2nd order in this paper. But we are afraid it will be very painful since according to Grozdanov et al. \cite{Grozdanov:3rd_hydro}, a total number of 68 new transport coefficients for the uncharged, nonconformal fluid will appear at the 3rd order. This number will reduce to 20 if one constrains the fluid into conformal regime. One fascinating question is will there be any relations like $4\ld_1+\ld_2=2\eta\tau_\pi$ exist in the 3rd or even higher derivative orders? Some recent frameworks for exploring high order hydrodynamics \cite{GaoJianHua,Heahl:eightfold_hydro_PRL,Heahl:eightfold_hydro_JHEP} may be helpful in this direction. Another choice for learning high order hydrodynamics may be at the linearized limit \cite{Buyanyan1,Buyanyan2,Buyanyan3,Buyanyan4}. But this framework may not answer the above question since it can not reach the coefficients like $\ld_{1,2}$ of 2nd order which relate with nonlinear viscous tensors.

Considering the discussions about the literatures on 2nd order strongly coupled hydrodynamics and the achievement that we have made in this paper, there are still some aspects valuable for future explorations. Firstly, one can use the Green-Kubo formula to calculate the thermal 2nd order coefficients for the background in this paper. Because of its inner structure, the original framework of BDE formalism of fluid/gravity correspondence is only able to extract $\ld_3$ and $\xi_3$ among the 8 thermodynamical coefficients. But as Ref.\cite{Finazzo:holo_nonconf_2nd} have shown us that the Green-Kubo formalism is good at extracting them. We are expecting to get at least $\kappa,~\kappa^*$ and $\xi_5$ not only because they are both from the 2-point correlation function hence relatively easier to calculate, but also these three coefficients form closed constraint equations \cite{Bhattacharyya:2nd_constraint}. Secondly, to calculate the entropy flux. Refs.\cite{Bigazzi:holo_nonconf_2nd_analytic,Kanitscheider:Dp_hydro,Finazzo:holo_nonconf_2nd} talking about the strongly coupled nonconformal relativistic fluid both do not mention the entropy flux. But this subject is reachable in the BDE formalism of fluid/gravity correspondence \cite{Bhattacharyya:entropy_flux} which is also a good aspect to explore. Thirdly, considering the nonconformal version that we have developed in \cite{Wu:fluid_gravity_D4_1st} and this paper, it is direct to calculate the 2nd order coefficients for the near-extremal black Dp-brane \cite{Kanitscheider:Dp_hydro,Mas:Dp_hydro_GreenKubo} to test the method of Ref.\cite{Kanitscheider:Dp_hydro}. Finally, it is interesting to add the smeared D0-brane charge into the compactified D4-brane \cite{Seki1304,Wuchao:D0D4_SS} to study the nonconformal fluid with a background vector charge. This framework can be seen as a nonconformal counterpart of \cite{Erdmenger:AdS5_U(1)charged,Banerjee:AdS5_U(1)charged} from a technical point of view. If adding a Chern-Simons term of D0-branes RR field, we may study the Chiral Vortical Effect for the nonconformal relativistic fluid in D0-D4 Sakai-Sugimoto model \cite{Wu:D0-D4_CVE}.

\section*{Acknowledgement}

We would like to thank Yu Lu for his great help on the calculation technique via computer. We also want to thank Yanyan Bu, Johanna Erdmenger, Zhang-Yu Nie for very helpful discussions on the second order calculation in the BDE formalism of fluid/gravity correspondence, and Shi Pu for the discussions on some physical aspects as well as the introduction to some literatures on relativistic hydrodynamics. C. Wu would like to thank the hospitality of the Wigner Research Center for Physics, Hungarian Academy of Sciences since his stay from November 5th 2016. This work is supported by the NSFC under Grant No. 11275213, and 11261130311(CRC 110 by DFG and NSFC), CAS key project KJCX2-EW-N01.

\appendix

\section{The dimensional reduction for the action of compactified D4-brane}

The total action for the compactified D4-brane background contains the bulk action, the Gibbons-Hawking boundary term and the counter term, which reads, in string frame as \cite{Bigazzi:dynamical_flavor_SS}
\begin{align}\label{10D total action string frame}
  S =& \frac1{2\kappa_{10}^2}\int d^{10}x \sqrt{-G^{(s)}} \left[ e^{-2\phi}\left(\mathcal R + 4(\nabla_{\hat M}\phi)^2 \right) - \frac{g_s^2}{2\cdot4!}F_4^2\right] \cr
  &- \frac1{\kappa_{10}^2} \int d^9x \sqrt{-H^{(s)}} e^{-2\phi}\mathcal K^{(s)} + \frac1{\kappa_{10}^2} \int d^9x \sqrt{-H^{(s)}} \frac5{2L}e^{-\frac73\phi},
\end{align}
where $2\kappa_{10}^2=(2\pi)^7g_s^2l_s^8$. Note $\phi$ here is the dilaton with zero vacuum expectation value, which is not same as in Ref.\cite{Bigazzi:dynamical_flavor_SS}. The first term of the above is the 10D bulk action with $G^{(s)}_{\hat M\hat N}$ the 10D metric in string frame. ``$\hat M,\hat N$" are the spacetime indices of 10D, $\mathcal R$ is the 10D Ricci scalar. The second term is the Gibbons-Hawking term with $H^{(s)}_{\hat M\hat N}$ the boundary metric and $\mathcal K^{(s)}=-H_{(s)}^{\hat M\hat N}\nabla_{\hat M}\mathbf n^{(s)}_{\hat N}$ the external curvature of 10D spacetime in string frame. $\mathbf n^{(s)}_{\hat M}\equiv\frac{\nabla_{\hat M}r}{\sqrt{G_{(s)}^{\hat N\hat P}\nabla_{\hat N}r\nabla_{\hat P}r}}$ is the 10D unit normal vector in string frame pointing to the direction of increasing $r$. The third term is the counter term.

Now we will reexpress the 10D total action of string frame into Einstein frame. For the bulk metric $G^{(s)}_{\hat M\hat N}$ we already know it will transfer to Einstein frame by $G^{(s)}_{\hat M\hat N}=e^{\frac{\phi}2}G_{\hat M\hat N}$ where $G_{\hat M\hat N}$ is the 10D bulk metric in Einstein frame. Thus the transformation rule for the unit norm $\mathbf n_{(s)}^{\hat M}$ is
\begin{align}
  \mathbf n^{(s)}_{\hat M} = \frac{\nabla_{\hat M}r}{\sqrt{G_{(s)}^{\hat N\hat P}\nabla_{\hat N}r\nabla_{\hat P}r}} = \frac{\nabla_{\hat M}r}{ \sqrt{ e^{-\frac{\phi}2} G^{\hat N\hat P} \nabla_{\hat N}r \nabla_{\hat P}r } } = e^{\frac{\phi}4} \mathbf n_{\hat M}.
\end{align}
Here $\mathbf n^{\hat M}$ is the 10D unit norm in Einstein frame. Then we have the transformation rule for $H^{(s)}_{\hat M\hat N}$ as
\begin{align}
  H^{(s)}_{\hat M\hat N} = G^{(s)}_{\hat M\hat N} - \mathbf n^{(s)}_{\hat M}\mathbf n^{(s)}_{\hat N} = e^{\frac{\phi}2} (G_{\hat M\hat N} - \mathbf n_{\hat M}\mathbf n_{\hat N}) = e^{\frac{\phi}2} H_{\hat M\hat N},
\end{align}
where $H_{\hat M\hat N}$ is the induced metric on a hyperplane at constant $r$ in the Einstein frame and its components can be read from
\begin{align}\label{ansatz for dimensional reduction of 9D boundary}
  ds^2 = e^{-\frac{10}3A} h_{MN}dx^M dx^N + e^{2A+8B}dy^2 + L^2 e^{2A-2B}d\Omega_4^2,
\end{align}
where $h_{MN}$ is the induced metric on a hyperplane at constant $r$ in the 5D reduced spacetime and the indices like ``$M,N$" are 5 dimensional. Eq.(\ref{ansatz for dimensional reduction of 9D boundary}) is actually the boundary of the following metric
\begin{align}\label{ansatz for dimensional reduction of 10D bulk metric}
  ds^2 = e^{-\frac{10}3A} g_{MN}dx^M dx^N + e^{2A+8B}dy^2 + L^2 e^{2A-2B}d\Omega_4^2,
\end{align}
which is the ansatz for dimensional reduction of the 10D bulk metric. The procedure to fix the coefficients in front of $A,B$ in Eq.(\ref{ansatz for dimensional reduction of 10D bulk metric}) can be found in Ref.\cite{Wu:fluid_gravity_D4_1st}. From (\ref{ansatz for dimensional reduction of 10D bulk metric}) one can see that $\hat M=\{M,y,\theta^a\}$ with $\theta^a$ the coordinates on the $S^4$.

Now we are ready to derive the transformation rule for the external curvature:
\begin{align}
  \mathcal K^{(s)} &= -H_{(s)}^{\hat M\hat N}\nabla_{\hat M}\mathbf n^{(s)}_{\hat N} = -e^{-\frac{\phi}2} H^{\hat M\hat N}\nabla_{\hat M}(e^{\frac{\phi}4} \mathbf n_{\hat N}) \cr
  &= -\left( e^{-\frac{\phi}4} H^{\hat M\hat N} \nabla_{\hat M} \mathbf n_{\hat N} + e^{-\frac{\phi}2} (\nabla_{\hat M}e^{\frac{\phi}4}) H^{\hat M\hat N}\mathbf n_{\hat N} \right) = e^{-\frac{\phi}4} \mathcal K.
\end{align}
Note that $H^{\hat M\hat N}\mathbf n_{\hat N}=0$. Here we define $\mathcal K$ is the 10D external curvature in Einstein frame. Using the fact that $\sqrt{-H^{(s)}}=e^{\frac94\phi}\sqrt{-H}$, one can reexpress the Gibbons-Hawking term into the Einstein frame as
\begin{align}
  S_{GH} &= - \frac1{\kappa_{10}^2} \int d^9x \sqrt{-H^{(s)}} e^{-2\phi}\mathcal K^{(s)} = - \frac1{\kappa_{10}^2} \int d^9x e^{\frac94\phi} \sqrt{-H} e^{-2\phi} e^{-\frac{\phi}4} \mathcal K \cr
  &= - \frac1{\kappa_{10}^2} \int d^9x \sqrt{-H} \mathcal K.
\end{align}
And the counter term turns out to be
\begin{align}
  S_{c.t.} &= \frac1{\kappa_{10}^2} \int d^9x \sqrt{-H^{(s)}} \frac5{2L}e^{-\frac73\phi} = \frac1{\kappa_{10}^2} \int d^9x e^{\frac94\phi} \sqrt{-H} \frac5{2L} e^{-\frac73\phi} \cr
  &= \frac1{\kappa_{10}^2} \int d^9x \sqrt{-H} \frac5{2L} e^{-\frac1{12}\phi}
\end{align}
in the Einstein frame. The details for getting the bulk action in Einstein frame can be found in standard textbooks so will be omitted here. Thus the total action in Einstein frame for the compactified D4-brane is
\begin{align}
  S &= \frac1{2\kappa_{10}^2} \int d^{10}x \sqrt{-G} \left[ \mathcal R - \frac12 (\del_{\hat M}\phi)^2 - \frac{g_s^2}{2\cdot4!} e^{\frac{\phi}2} F_4^2 \right] - \frac1{\kappa_{10}^2} \int d^9x \sqrt{-H} \mathcal K \cr
  &+ \frac1{\kappa_{10}^2} \int d^9x \sqrt{-H} \frac5{2L} e^{-\frac1{12}\phi}.
\end{align}
Note that with the appearance of $L$ in the denominator, the counter term has the same dimension with the Gibbons-Hawking term.

Now we will use Eqs.(\ref{ansatz for dimensional reduction of 9D boundary}) and (\ref{ansatz for dimensional reduction of 10D bulk metric}) to reduce the last total action into 5D form. The reducing procedure for the bulk action can be found in Ref.\cite{Wu:fluid_gravity_D4_1st} and will be omit here, we mainly care about $S_{GH}$ and $S_{c.t.}$. Firstly, we have from Eq.(\ref{ansatz for dimensional reduction of 9D boundary}) that $\sqrt{-H}=L^4e^{-\frac53A}\sqrt{-h}\sqrt{\gamma_4}$, where $\gamma_4$ is the determinant of $\gamma_{ab}$, the metric on the unit 4-sphere. From the definition of $\mathbf n_{\hat M}$, one can see that the components of the 10D unit norm in the directions of $y$ and $\theta^a$ are both 0 thus one has $\mathbf n_{\hat M}=(\mathbf n_M,0,\boldsymbol0)$ with
\begin{align}
  \mathbf n_M &= \frac{\nabla_{M}r}{\sqrt{G^{\hat N\hat P}\nabla_{\hat N}r\nabla_{\hat P}r}} = \frac{\nabla_M r}{ \sqrt{ { G^{NP} \nabla_N r \nabla_P r} } } = \frac{\nabla_M r}{ \sqrt{ e^{\frac{10}3A}g^{NP} \nabla_N r \nabla_P r} }  \cr
  & = e^{-\frac53A}\frac{\nabla_M r}{ \sqrt{ g^{NP} \nabla_N r \nabla_P r} } = e^{-\frac53A} n_M,
\end{align}
where we use $G_{MN}=e^{-\frac{10}3A}g_{MN}$ from Eq.(\ref{ansatz for dimensional reduction of 10D bulk metric}). Note also that in fact
\begin{align}
  G^{\hat N\hat P}\nabla_{\hat N}r \nabla_{\hat P}r = G^{NP} \nabla_N r \nabla_P r,
\end{align}
since one has $\nabla_yr = \nabla_{\theta^a}r =0$. We also define the unit norm of 5D as $n_M\equiv\frac{\nabla_M r}{\sqrt{g^{NP} \nabla_Nr \nabla_Pr}}$. So the external curvature of 10D can be reduced to
\begin{align}
  \mathcal K &= -H^{\hat M\hat N}\nabla_{\hat M} \mathbf n_{\hat N} = -H^{MN} \nabla_M \mathbf n_N = -e^{\frac{10}3A} h^{MN} \nabla_M (e^{-\frac53A}n_N) \cr
  &= -e^{\frac53A} h^{MN} \nabla_Mn_N = e^{\frac53A} K,
\end{align}
where $K=-h^{MN}\nabla_Mn_N$ is the external curvature in 5D. We have used the fact that $\nabla_y\mathbf n_M=\nabla_{\theta^a}\mathbf n_M=0$ since $\mathbf n_M$ will only depend on $r$. Also note that $h^{MN}n_N=0$. So finally we can reduce $S_{GH}$ as
\begin{align}
  S_{GH} &= - \frac1{\kappa_{10}^2} \int d^9x \sqrt{-H} \mathcal K = - \frac1{\kappa_{10}^2} \int d^4x \int dy \int d^4\theta \sqrt{\gamma_4} L^4e^{-\frac53A}\sqrt{-h} e^{\frac53A} K \cr
  &= - \frac{L^4\Omega_4\beta_y}{\kappa_{10}^2}\int d^4x \sqrt{-h}K = - \frac1{\kappa_5^2}\int d^4x \sqrt{-h}K,
\end{align}
where $\int dy=\beta_y$, $\int d^4\theta \sqrt{\gamma_4}=\Omega_4$ and we have defined
\begin{align}
  \frac1{\kappa_5^2} \equiv \frac{L^4\Omega_4\beta_y}{\kappa_{10}^2}.
\end{align}
Then for the counter term, we have
\begin{align}
  S_{c.t.} &= \frac1{\kappa_{10}^2} \int d^9x \sqrt{-H} \frac5{2L} e^{-\frac1{12}\phi} = \frac1{\kappa_{10}^2} \int d^4x dy d^4\theta L^4 e^{-\frac53A} \sqrt{-h} \sqrt{\gamma_4} \frac5{2L} e^{-\frac\phi{12}} \cr
                &= \frac{L^4\Omega_4\beta_y}{\kappa_{10}^2} \int d^4x \sqrt{-h} \frac5{2L} e^{-\frac53A-\frac\phi{12}} = \frac1{\kappa_5^2} \int d^4x \sqrt{-h} \frac5{2L} e^{-\frac53A-\frac\phi{12}}.
\end{align}
And the details for the dimensional reduction of the bulk action can be found in Ref.\cite{Wu:fluid_gravity_D4_1st}. So the total action of the 5D reduced system is
\begin{align}
  S &= \frac1{2\kappa_5^2}\int d^5x\sqrt{-g}\left[ R - \frac12(\p\phi)^2 - \frac{40}3(\p A)^2 - 20(\p B)^2 - V(\phi,A,B) \right] \cr
  &- \frac1{\kappa_5^2}\int d^4x \sqrt{-h}K + \frac1{\kappa_5^2}\int d^4x \sqrt{-h} \frac5{2L} e^{-\frac53A-\frac1{12}\phi},
\end{align}
where
\begin{align}
  V(\phi,A,B) = \frac{Q_4^2}{2L^8} e^{\frac\phi2 - \frac{34}3A + 8B} - \frac{12}{L^2} e^{-\frac{16}3A+2B}.
\end{align}

\end{document}